# The Lunar Geophysical Network Landing Sites Science Rationale

Heidi Fuqua Haviland[1] [2], Renee C. Weber[2], Clive R. Neal[3], Philippe Lognonné[4], Raphaël F. Garcia[5], Nicholas Schmerr[6], Seiichi Nagihara[7], Robert Grimm[8], Douglas G. Currie[6], Simone Dell'Agnello[9], Thomas R. Watters[10], Mark P. Panning[11], Catherine L. Johnson[12] [13], Ryuhei Yamada[14], Martin Knapmeyer[15], Lillian R. Ostrach[16] [17], Taichi Kawamura[4], Noah Petro[18], Paul M. Bremner[2]

## Abstract

The Lunar Geophysical Network (LGN) mission is proposed to land on the Moon in 2030 and deploy packages at four locations to enable geophysical measurements for 6-10 years. Returning to the lunar surface with a long-lived geophysical network is a key next step to advance lunar and planetary science. LGN will greatly expand our primarily Apollo-based knowledge of the deep lunar interior by identifying and characterizing mantle melt layers, as well as core size and state. To meet the mission objectives, the instrument suite provides complementary seismic, geodetic, heat flow, and electromagnetic observations. We discuss the network landing site requirements and provide example sites that meet these requirements. Landing site selection will continue to be optimized throughout the formulation of this mission. Possible sites include the P-5 region within the Procellarum KREEP Terrane (PKT; (lat:15°; long:-35°), Schickard Basin (lat:-44.3°; long:-55.1°), Crisium Basin (lat:18.5°; long:61.8°), and the farside Korolev Basin (lat:-2.4°; long:-159.3°). Network optimization considers the best locations to observe seismic core phases, e.g., ScS and PKP. Ray path density and proximity to young fault scarps are also analyzed to provide increased opportunities for seismic observations. Geodetic constraints require the network to have at least three nearside stations at maximum limb distances. Heat flow and electromagnetic measurements should be obtained away from terrane boundaries and from magnetic anomalies at locations representative of global trends. An in-depth case study is

[1] Corresponding author: heidi.haviland@nasa.gov

[2] NASA Marshall Space Flight Center, Heliophysics and Planetary Science Branch, 320 Sparkman Ave, NSSTC/2070, Huntsville, AL 35820

[3] University of Notre Dame, Dept. Civil & Env. Eng. & Earth Sciences

[4] Université de Paris, Institut de physique du globe de Paris, CNRS, F75005 Paris.

[5] ISAE-SUPAERO, 10 ave E. Belin F-31400 Toulouse, France

[6] University of Maryland at College Park

[7] Texas Tech University

[8] Southwest Research Institute

[9] Istituto Nazionale di Fisica Nucleare, Laboratori Nazionali di Frascati (INFN-LNF), Frascati, Italy

[10] Center for Earth and Planetary Studies of the National Air and Space Museum, Smithsonian Institution, Washington, DC 20560, USA

[11] NASA Jet Propulsion Lab

[12] Department of Earth, Ocean and Atmospheric Sciences, University of British Columbia, Vancouver, British Columbia, Canada

[13] Planetary Science Institute, Tucson, AZ, USA

[14] Aizu University, Japan

[15] DRL, Berlin, Germany

[16] U.S. Geological Survey, Astrogeology Science Center

[17] "This draft manuscript is distributed solely for purposes of scientific peer review. Its content is deliberative and predecisional, so it must not be disclosed or released by reviewers. Because the manuscript has not yet been approved for publication by the U.S. Geological Survey (USGS), it does not represent any official USGS finding or policy."

[18] NASA Goddard Space Flight Center, Greenbelt, MD 20771

provided for Crisium. In addition, we discuss the consequences for scientific return of less than optimal locations or number of stations.

## 1. INTRODUCTION

Understanding the interior properties of the Moon, including the size and state of the core, is of utmost importance to the future of lunar and planetary science. While a wealth of information has been gleaned from Apollo-era investigations and subsequent orbiters, gaps remain in our current understanding (Neal et al., 2020). A long-lived next-generation network of surface geophysical stations, the Lunar Geophysical Network (LGN), will provide simultaneous multipoint geophysical observations across four complementary disciplines: seismology, geodesy, heat flow, and electromagnetics from around the Moon. Together these observations will unlock key outstanding issues regarding the lunar interior including the existence of, size, and state of the inner core; the presence of a deep mantle partial melt layer; mantle thermal state; and composition including lateral and vertical heterogeneity.

The purpose of this paper is to discuss the science driving the LGN mission landing sites. In particular, we discuss a network comprised of four sites with seismic, laser ranging (on the nearside), heat flow, and magnetotelluric measurements made at each. We provide an in-depth discussion of a possible site in Mare Crisium, as a case study for future mission planning methodology for locating where within a landing site is best for the LGN mission as a whole. We begin with example landing sites and note that these will continue to evolve through the formulation of the mission. We also note that the LGN mission requires continuous measurements to be made throughout multiple lunar day-night cycles.

In addition to LGN, there are several exciting upcoming opportunities for missions to the lunar surface, including short duration (< 14 Earth days, limited to operate within the daytime without night survival) Commercial Lunar Payload Services (CLPS) missions, the future NASA Artemis program providing a human presence at the lunar surface, and international missions such as the European Lunar Geophysical Observatory (ELGO) (Garcia et al, 2020). In this paper, we focus on the four LGN stations as primary nodes to achieve the mission science objectives with a launch in 2030. Other opportunities may help improve instrument performance, risk mitigation, and to provide additional nodes in support of the LGN network, however, discussion of these opportunities is beyond the scope of this paper. We first provide a description of the LGN mission, including science questions, objectives, and traceability; an overview of the communications architecture required by this mission; and required instrument performances (Section 2). We highlight four potential landing sites, review the implications of optimizing the locations of these primary network nodes as well as the scientific consequences and priorities of needing to remove one or more landing sites. In addition, we detail the seismic, geodetic, heat flow, and electromagnetic science requirements for the network (Section 3). We then provide a detailed study on how to locate a surface site within Mare Crisium as a case study (Section 4). Lastly, we conclude by highlighting next steps for this mission in determining landing sites (Section 5), and conclude. The reader is directed to Table S1 for a list of all acronyms used.

## 2. THE LUNAR GEOPHYSICAL NETWORK MISSION OVERVIEW

The goal of the LGN mission is to understand the evolution of terrestrial planets, from their initial stages of formation, differentiation, and subsequent persistence (or lack) of internal dynamics into the present (Neal et al., 2020). Terrestrial planets all share a common structural framework (e.g., crust, mantle, core) which is developed very shortly after formation and that determines subsequent evolution (e.g., McCulloch, 1987; Elkins-Tanton et al., 2003, 2011; Elkins-Tanton, 2008; Brown and Elkins-Tanton, 2009; Charlier et al., 2013; Maurice et al., 2017; Ikoma et al., 2018). The Moon is a natural target for this type of geophysical network mission as it presents an opportunity to study an internal heat engine that waned early in planetary evolution, and thereby enabled preservation of the initial magma ocean differentiation event (Smith et al., 1970; Wood et al., 1970). Such information has been lost on Earth due to crustal recycling and weathering of our most ancient rocks, and also on Mars and Venus due to their larger sizes and heat engines, producing prolonged volcanic activity and resurfacing that is thought to have continued to the present day (e.g., Hartmann et al., 1999; Stofan et al., 2016; Filiberto et al., 2020). The lunar initial differentiation model is supported by analyses of returned Apollo basaltic samples, which are consistent with derivation from a source comprised of cumulates that crystallized from an initial magma ocean and subsequently underwent an overturn event (Taylor and Jakes, 1974; Snyder et al., 1992; 1997).

A LGN mission should be (1) "better than Apollo" (see Section 2.1, below), (2) permit a global distribution of stations, including the farside, and (3) allow for redundancy as part of the baseline mission. Each lander should contain a sensitive broadband seismometer, laser retroreflectors, heat flow probes, and electromagnetic sounders. The landers should be long-lived (>6 years, with a goal of 10 years) to maximize science and allow other nodes to be added by international and commercial partners during the lifetime of the mission, thus increasing the fidelity and value of the data obtained.

The baseline LGN mission architecture requires an orbiter to serve as the primary communication relay for the farside lander. This concept requires mission operations to manage command and control of the orbiter and four landers. The mission operations team should primarily utilize direct to Earth communication links to talk to all landers on the nearside of the Moon. Operations will have the option to communicate with the nearside landers via the orbiter relay communication links for redundancy.

### 2.1 Science Questions and Objectives

Our first look into the Moon's interior came from the Apollo Lunar Surface Experiment Packages (ALSEP) that deployed surface magnetometers, placed laser retroreflector arrays, installed seismometers that detected moonquakes and meteorite impacts, and took heat flow measurements – key geophysical information that has advanced our knowledge of the Moon's internal structure, evolution, and dynamics. However, it is now evident that due in part to the relatively narrow geographical extent of the Apollo passive seismic network, our understanding of the lunar interior, and especially the deep interior and core, remains incomplete (Figure 1). For example, garnet has been hypothesized to exist below ~500 km within the Moon (e.g., Anderson, 1975; Hood, 1986; Hood and Jones, 1987; Neal, 2001). However, interpretations based on the limited seismic data are ambiguous. For example, Nakamura et al. (1974) and Nakamura (1983) suggested that higher velocities in the middle mantle (>500 km) could be indicative of an increased proportion of Mg-rich olivine. This has implications for the bulk composition of the Moon, especially for $Al_2O_3$. For example, GRAIL data have been used to constrain the crustal contribution to bulk $Al_2O_3$ of 1.7-2.1 wt.% (Wieczorek et al., 2013) and conclude that the bulk $Al_2O_3$ content of the Moon is lower than the 6.1 wt.% of the Taylor Bulk Moon composition (Taylor, 1982). However, Wieczorek et al. (2013) note that $Al_2O_3$ in the mantle derived from Apollo seismic data have a broad range (2.0-6.7 wt.%) for the upper and lower mantle. Khan et al. (2006a) demonstrated that no mid-mantle velocity jump was required to fit the Apollo seismic data with a homogeneous mantle bulk composition. Stable phases have been calculated in compositional models for the lunar interior, along with their respective seismic properties. The results suggest that garnet would be a minor stable phase in the middle mantle of the Moon between 270 and 500 km, but would be a significant phase in the lower lunar mantle between 500 and 1262 km (e.g., Khan et al. 2006a; Kuskov et al., 2014; Kuskov et al., 2019 and references therein). The existing seismic data are insufficient to be used to discriminate between these models and yield accurate estimates of lunar bulk composition.

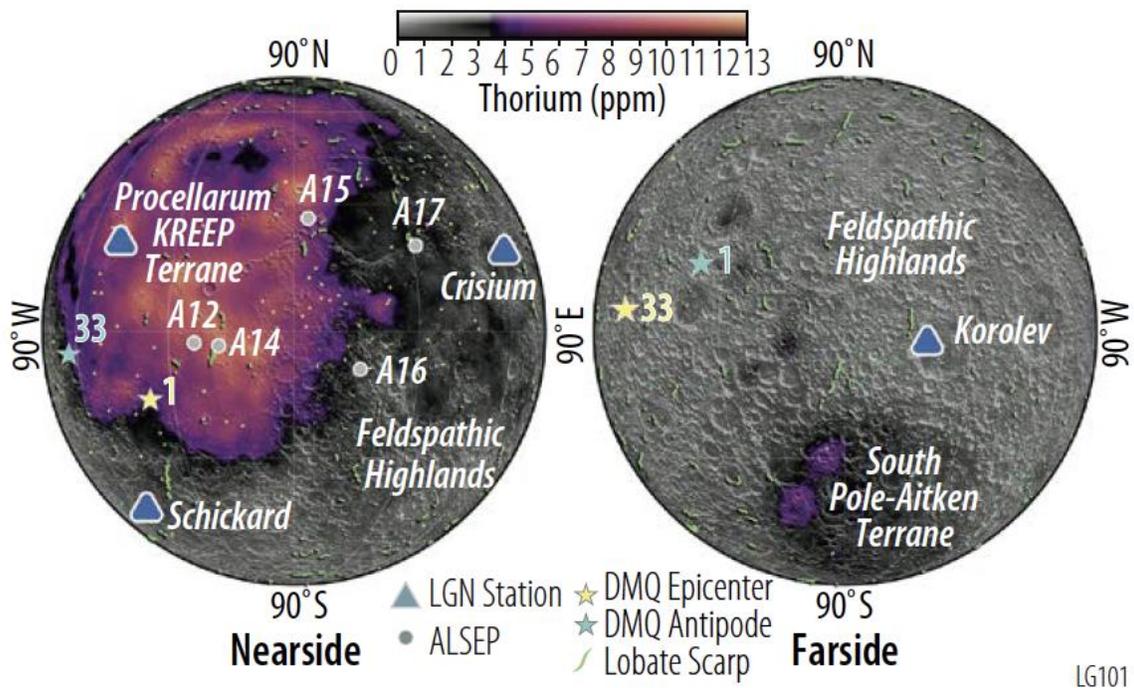

Figure 1. Example Lunar Geophysical Network station locations (blue triangles) compared to Apollo (grey circles). LGN stations will be placed across major lunar terranes and enable new interrogation of the deep lunar interior and tectonic evolution. The proposed LGN stations are positioned to take advantage of 1) possible recent lobate scarp seismicity (green lines, Watters et al., 2019) and 2) known deep moonquake (DMQ) clusters and their antipodes (yellow and cyan stars, respectively). The two most active nearside and farside deep moonquake clusters (A01 and A33) are highlighted. DMQ cluster epicenters, each exhibiting characteristically repeating

waveforms between 550-1420 km in depth, are thought to be caused by the tidal forces of the Earth-Moon system at 27 to 29 day intervals (Weber et al., 2009). Their antipodes, where core transmitted phases would be focused, are located on the opposite side of the Moon (Nakamura, 2005). The LGN network is designed to geophysically interrogate the internal structure, temperature, composition, and tectonics/seismicity both in the Feldspathic Highlands (unsampled by Apollo) and within the Procellarum KREEP terrane, outlined by the Lunar Prospector thorium abundance (Lawrence et al., 2000). Shaded surface relief is derived from LOLA topography (Smith et al., 2017).

Additionally, the identification of different lunar surface terranes from new global surface compositional data have produced a shift in our understanding of the global evolution model of the Moon (Jolliff et al., 2000; Laneuville et al., 2018). These data demonstrate that there is a fundamental limitation to the ALSEP geophysical datasets, as all were collected in or very near one anomalous region, the Procellarum KREEP (Potassium, Rare Earth Elements, Phosphorus) Terrane (PKT) (Figure 1). Subsequent orbital missions have expanded the global geophysical picture of the interior (e.g., Lunar Prospector (LP), Gravity Recovery And Interior Laboratory (GRAIL), Kaguya, Lunar Reconnaissance Orbiter (LRO), etc.), but only a landed long-lived geophysical network can address the significant questions that remain unanswered by Apollo:

- How does the overall composition and structure of the Moon inform us about initial differentiation of terrestrial planets?
- What is the state, structure, and composition of the mantle and is it consistent with the lunar magma ocean hypothesis (or are there resolvable discontinuities)?
- What range of bulk compositional models for the crust, mantle, and core are compatible with combined LGN and previous datasets?
- What is the present heat budget and how could the Moon experience magmatism for >3 b.y.?
- What is the crust and mantle heterogeneity within and between different terranes?
- Based on a constrained size, state and composition, how did the lunar core form and could it have supported a global magnetosphere (as indicated by sample analyses – e.g., Weiss and Tikoo, 2014; Mighani et al., 2020)?

New geophysical data are needed to address these questions and add greater fidelity to datasets already obtained. The Apollo seismic network possessed a narrow aperture and was emplaced on the edge of and within a special crustal terrane, but has necessarily been used to extrapolate global interior properties. For example, the GRAIL gravity data that inform crustal thickness are constrained by Apollo seismic data (Khan and Mosegaard, 2002; Lognonne et al., 2003; Chenet et al., 2006), but the fidelity across the lunar surface is poor due to the narrow aperture of the Apollo passive seismic network and localization in the thinned and likely anomalous nearside crust of the PKT (e.g., Hood, 1986). Another example is the large discrepancies among the size (and nature) of the lunar core defined by seismic, lunar laser retroreflector (LLR), and GRAIL data, which reflect the lack of fidelity in Apollo seismic and ongoing LLR data (e.g., Williams et al., 2014). Therefore, the fundamental purpose of the LGN mission is to broadly distribute landers with seismometers, heat flow probes, Next Generation Lunar Retroreflectors (NGLR), and electromagnetic sounders around the Moon, including on the farside, that are placed well within the boundaries of varied lunar terranes from which global properties can be extracted A similar analogy can be drawn for the Earth: unless geophysical information can be drawn across a representative number of terranes on our own planet (e.g., oceanic plates, continents, cratons, plate margins, etc.), it would have been impossible to formulate an accurate geophysical picture of the different internal processes at work within the Earth. Indeed, it was shortly after geophysical exploration of the oceanic plates commenced that the paradigm shift to plate tectonics took place. Terrestrial seismology also enables elastic constraints to be related back to mineral properties within the deep interior structure of the Earth. Thus, a global network of geophysical stations will enable LGN to fully interrogate the deep interior of the Moon, more accurately locate hypocenters of large moonquakes and impacts, and constrain crustal thickness variations across a wide range of lunar terranes – none of which are possible with the Apollo data.

Electrical conductivity can provide constraints on two key components of the interior, iron and water. These constituents dominate because they are the strongest contributors to lattice defects that foster mobile electrical charges. The Arrhenius relationships favor iron as the highest conductivity at temperatures >1000-1500 K and water at lower temperatures (see Verhoeven and Vacher, 2018, for a review). Because they have different activation energies, the relative contributions of iron and water can be separated along a geotherm. While simultaneous inversion of temperature and composition from electrical conductivity is possible, a more robust result is obtained by combining these data with heat flow, which provides an independent estimate of the geotherm. The joint heat flow and electrical-conductivity constraints on temperature and iron/water content can be combined with seismological estimates of temperature and bulk mineralogy for further refinement of the state of the Moon's interior.

The LGN mission will allow more intricate questions to be addressed that have resulted from previous work, such as:

- Do shallow moonquakes represent movement along thrust faults (e.g., Watters et al., 2019)?
- Do moonquakes present a threat to future human infrastructure (Oberst and Nakamura, 1992)?
- Do deep moonquakes occur on the farside of the Moon (Nakamura et al., 1982; Nakamura, 2005)?
- Is the appearance that the farside is aseismic an artifact of network aperture, interior structure, or is it linked to the presence of maria on the nearside only (e.g., Watters & Johnson 2010; Qin et al., 2012; Laneuville et al., 2018)?
- What is the mechanism for triggering deep moonquakes (Weber et al., 2009; Kawamura et al., 2017)?
- Are there global discontinuities in the mantle and do they relate to the lunar magma ocean (Nakamura et al., 1982; Lognonné et al., 2003)?
- Do different lunar terranes have unique heat flow budgets and what does this imply about the bulk geochemical composition of the Moon (Laneuville et al., 2018)?
- What is the lateral/vertical structure and composition as revealed by electrical conductivity (Hood et al., 1982; Grimm, 2013; Khan et al., 2006b, 2014; Shimizu et al., 2013)?
- What are the differences between near and farside hemispheres? (e.g., Jolliff et al., 2000) Are differences observed at the surface manifest in the interior and if so, how? (e.g., Wieczorek and Phillips, 2000)

- What is the nature of the presumed "partial melt layer" on top of the core mantle boundary, and why does it exist? (e.g., Elkins-Tanton et al., 2002, Weber et al 2011, Khan et al 2014)
- Can brittle failure occur at depths where the mantle should be hot? (e.g., Khan and Mosegaard, 2002; Nimmo et al., 2012; Khan et al., 2013; Kawamura et al., 2017)

Several lines of evidence indicate that the lunar mantle likely preserves a vertical compositional gradient and at least one internal discontinuity. A whole-moon magma ocean seems most consistent with rapid accretion following the giant impact. Therefore, the boundary between the first two generations of cumulates (olivine- vs pyroxene-dominated) would lie at a depth between 450 km (Synder et al., 1992) and 700 km (Elkins-Tanton et al., 2011). Mantle overturn would erase this boundary. Nonetheless, a contrast reappears at 700-km depth in the model of Elkins-Tanton and colleagues. Basaltic mare eruptions represent a later phase of magmatism whose source depths increased from ~150 to >500 km over the interval 4.3 to 1.2 Ga (e.g., Longhi et al., 1974; Walker et al., 1975; Longhi, 1992, 1993; Barr and Grove, 2013; Hiesinger et al., 2011a), probably due to global cooling and thickening of the lithosphere (e.g., Snyder et al., 1992). A compositional contrast would be left at the base of the mare-basalt source region; however, it is not required by the geophysics (Khan et al., 2006a). Again, mantle overturn may have disrupted initial layering below the floatation crust, but the upper few hundred km may still hold compositionally distinct materials, including sources of the Mg-suite.

The discovery of the compositional asymmetry of the Moon (Jolliff et al., 2000; and companion papers) calls into question whether the derived dynamics and petrological structure are truly global or are instead regional specific to that part of the nearside sampled by Apollo. The gamma-ray instrument on Lunar Prospector (LP) revealed that thorium is concentrated on the northwestern nearside of the Moon and, by inference, other incompatible elements that make up KREEP (potassium, rare-earth elements, and phosphorus). The anomalous region encompasses Oceanus Procellarum, Mare Imbrium, and adjacent mare and highlands and so was named the Procellarum KREEP Terrane, or PKT. The rest of the Moon is classified as Feldspathic Highlands Terrane (FHT), with the exception of the South Pole Aitken Basin (SPA). While comprising only 1/6 of the Moon's surface, PKT contains nearly two-thirds of the maria area (as well as the youngest mare basalt fields – Hiesinger et al., 2011a), and therefore, strongly suggests a connection between KREEP and melting of mare source regions (Wieczorek and Phillips, 2000). Korotev (2000) linked certain mafic impact-melt breccias (i.e., "Low-K Fra Mauro" or "LKFM"), and ultimately the entire Mg-suite, to KREEP-rich lower crust and upper mantle materials that are thus uniquely associated with PKT.

The mechanism by which "urKREEP"--the putative source region for KREEP--was geographically concentrated is uncertain (see Shearer et al., 2006, for a summary). Nonetheless, the distribution of residual urKREEP influences the temperature of the crust and upper mantle, which in turn has implications for the depth and duration of melting as well as topography and gravity. Wieczorek and Phillips (2000) modeled the thermal evolution of the Moon following emplacement of concentrated, subcrustal urKREEP. They successfully reproduced the differences in heat flow measured at the Apollo 15 and 17 sites, and the longevity and depth of melting are well matched to mare basalts. However, such a large thermal anomaly would produce large gravity or topography anomalies that are not observed. Instead, Grimm (2013) found that a similar thermal model in a lower concentration of urKREEP partitioned throughout the crust matched heat flow but damped the mantle thermal anomaly, thus satisfying gravity and topography. While melting depths still exceed 400 km, the duration of melting is reduced compared to the mantle-heating case. Finally, the upper 120 km crystallized as a stratified combination of anorthite, oxides, and diverse pyroxenes (Elkins-Tanton et al., 2011; Snyder et al.; 1992).

## 2.2 Science Traceability

LGN's primary goal is to *understand the initial stages of terrestrial planet evolution*. The instruments selected for LGN should have the capabilities required to make the measurements that enable its investigations and objectives, and ultimately answer the primary goal. Therefore, to achieve this goal, four ***Objectives*** explored through six ***Investigations*** have been identified.

**Objectives**:
• Evaluate the interior structure and dynamics of the Moon.
• Constrain the interior and bulk composition of the Moon.
• Delineate the vertical and lateral heterogeneities within the interior of the Moon as they relate to surface features and terranes.
• Evaluate the current seismo-tectonic activity of the Moon (Kumar et al., 2016, 2019; Watters et al., 2015, 2019).

**Investigations**:
• Determine the size and state of the core, and infer its composition, including the potential for lighter alloying elements, (building on the work of Williams et al., 2001, Weber et al., 2011 and Garcia et al., 2011).
• Determine the state and chemical/physical stratification of the lunar mantle (is there garnet in the lower mantle; e.g., Neal, 2001; Khan et al., 2006a; Kuskov et al., 2001?).
• Determine the thickness of the lunar crust and characterize its vertical and lateral variability (refining and adding fidelity to the GRAIL results of Wieczorek et al., 2013 and multiplying crustal receiver function analysis, Vinnick et al. 2001).
• Determine the thermal state of the lunar interior and elucidate the workings of the planetary heat engine.
• Monitor impacts on the lunar surface as an aid to exploring the lunar interior.
• Characterize the seismo-tectonic properties at the lunar surface in support of future human infrastructure (e.g., Oberst and Nakamura, 1992; Ortiz et al 2006; Banks et al, 2020).

## 2.3 Instrument Performance Requirements

### 2.3.1 Seismic

Despite being the first digital-feedback seismometer, the Apollo seismometers suffered from their low-resolution acquisition system (10 bit), which was not able to register the instrument self-noise on the Apollo peaked long-period (LP) vertical axis. Although the flat mode was broadband, it was not sensitive enough to capture the low magnitude deep moonquakes and most of the Apollo observations were therefore made with the narrow band peaked mode, with a peak sensitivity of about 5 x $10^{-11}$ m in displacement or 5 x $10^{-10}$ m/s$^2$ in acceleration at a period of 2 sec (see Lammlein et al., 1974 or the electronic supplement of Nunn et al., 2020 for instrument response curves). Even in this mode the digitization noise remains the main limitation for most of the signals (Figure 2a). Many deep moonquakes (DMQ) were therefore recorded with peak-to-peak amplitudes of only a few bits (Bulow et al., 2005, Lognonné & Johnson, 2015, Nunn et al., 2020). This leads to partial capture of seismic phases by the lunar seismograms, especially for the core phases, which are expected to be about 10-20 times smaller than the direct P and S phases (Figure 2b) and which can be revealed only through stacks (Garcia et al., 2011; Weber et al., 2011). Importantly for future lunar seismic experiments, modeling suggests that the lunar seismic noise may be at least a factor 10 below the Apollo peaked mode resolution (Lognonné et al., 2009), which implies that more events will be observed with greater instrument sensitivity.

Following the success of the Seismic Experiment for Interior Structure (SEIS) instrument on InSight (Lognonné et al., 2019, 2020; Giardini et al., 2020) and ongoing development of space qualification for seismic instrumentation (Nunn et al., 2019, Erwin et al., 2020), a sensitivity requirement ten times better than the Apollo long-period sensor can be conservatively achieved for a Very Broad Band (VBB) LGN instrument with a performance level of about 3.5 x $10^{-11}$ m/s$^2$/Hz$^{1/2}$ in the long-period bandwidth, 0.01-1 Hz frequency range. This sensitivity will allow the detection of core phases (Yamada et al., 2013) for deep interior seismic structure, as well as detection of low-mass meteorite impacts for crustal structure made in conjunction with impact flash monitoring (Yamada et al., 2011). At this sensitivity level, the lander itself will be a source of noise (Panning et al., 2020) and requires active monitoring made with a Short-Period (SP) seismometer comparable to the SEIS-SP (Lognonné et al., 2019), such as the Silicon Seismic Package (SSP). An additional SP seismometer will be also deployed on the VBB-LGN sensor assembly. With some expected improvement in an SP-like seismometer when tuned for lunar operation (Nunn et al., 2019), a noise floor comparable to (or within a factor of a few above the) Apollo peaked-mode sensitivity can be achieved. The addition of this SP sensor provides some redundancy for the VBB sensor, as well as improving capabilities via combination of six seismic sensors to obtain translational and rotational information (Fayon et al., 2018, Sollberger et al., 2016) of the short-period wavefield generated by thermal moonquakes.

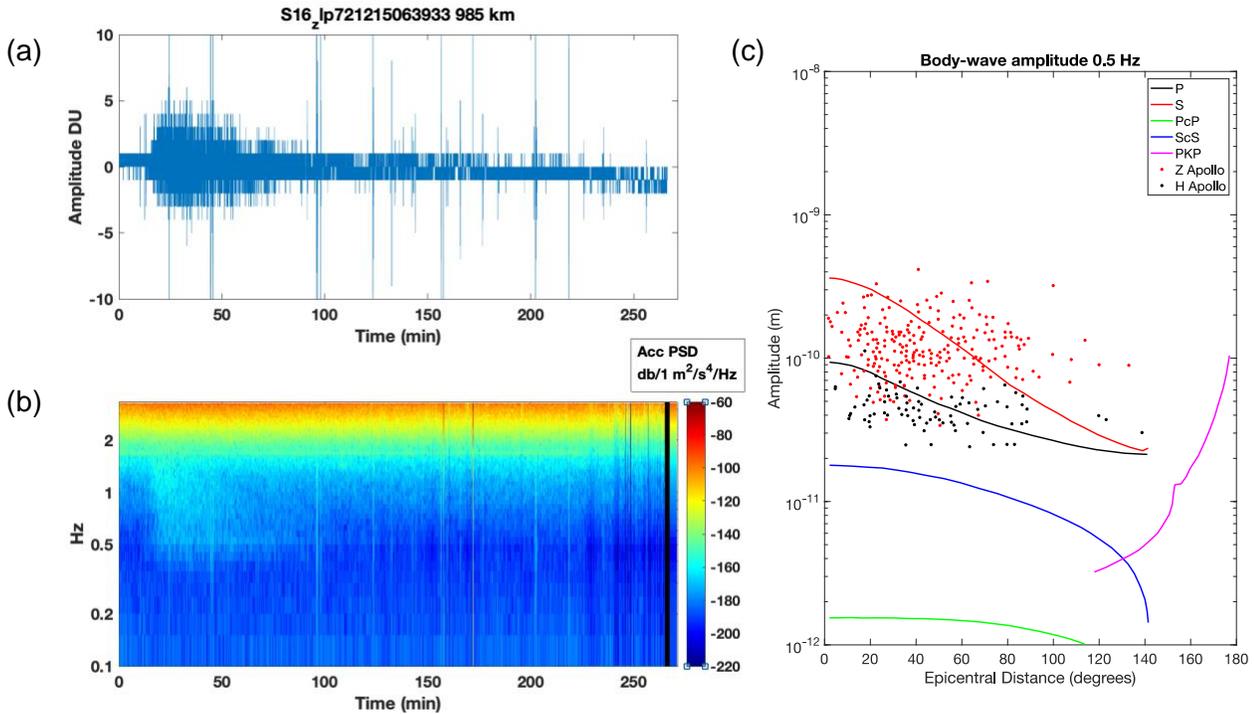

**Figure 2.** (a) The impact of the Apollo 16 ascent vehicle recorded on the Apollo 16 long-period seismometer. The digitization of the instrument is clearly visible as integer steps in the waveform. (b) A spectrogram, in acceleration power spectral amplitude, shows the spectral content and limitation of the 10-bit acquisition noise, shown in the final section of the spectrogram, after the black line. This 10-bit dynamic range also hinders the identification of seismic arrivals in the frequency domain. (c) LGN's amplitude sensitivity will be a factor of 10 better than Apollo as well as capturing a larger epicentral distance, enabling core phases to be resolved (e.g., core-reflected shear wave, ScS, core-traversing P-wave, PKP, core-reflected P-wave, PcP) on single records, in contrast to Apollo, which detected only direct P and S phases (Yamada et al., 2013). Red and black points are P and S phases observed from Apollo records, while the expected core phases were all below the instrument resolution and required stacking to detect. Theoretical amplitude of body waves (color lines) are shown for deep moonquakes as a function of epicentral distance.

### 2.3.2 Geodetic

Earth-based Lunar Laser Ranging Observatories (LLRO) were first developed in the 1960s to send short laser pulses toward the Apollo retroreflector arrays on the lunar surface where the Cube Corner Retroreflectors (CCRs) reflected the laser pulses directly back to the LLRO. By highly accurate timing of light travel time, the distance to the Apollo retroreflector array was determined to centimeter scale, that is, to better than one part in $10^{12}$. LLR has continued over the past 50 years and the analyses of these data have illuminated key details within lunar geophysics, including detecting: the existence and shape of the liquid core, an oblate spheroid with the short axis parallel to the rotation axis of the Moon (Williams et al., 2009); the $k_2$ Love number and tidal Qs (Williams and Boggs, 2015); Core Mantle Boundary (CMB) dissipation (Williams et al. 2001; 2014); active or geologically recent stimulation of rotation normal modes (Eckhardt 1993, Newhall and Williams, 1997, Chapront et al., 1999, Rambaux and Williams, 2011), improving lunar cartographic networks (with submeter accuracy at the five Apollo reflectors (Williams et al., 1996; 2013; Wagner 2017), as well as by providing a check on altimetry from orbit (Fok 2009). In addition, LLR provides the best tests of general relativity and gravitation, improving observations of the strong and weak equivalence principles (Williams et al 2012; Williams et al 2004) including providing the only test that measures properties of the gravitational binding energy and the nonlinear effects of gravitational self-interaction (Hofmann & Muller, 2018), and providing the best limits on the temporal change in gravitational constant (Williams et al 2004).

Due to the combination of the design of the Apollo retroreflector arrays and the lunar librations, the uncertainty or jitter in a single range measurement is about 67 mm. The use of the large single solid CCR incorporated in the Next Generation Lunar Retroreflector (NGLR) will improve the normal point accuracy by a factor of 10 or more leading to the concomitant improvement in the accuracy of the science results for the LGN NGLR array, in particular, these tests include the weak equivalence principle and the inertial properties of gravitational energy, the strong equivalence principle and the spatial and temporal variation of the gravitational constant. To obtain the full accuracy supported by the NGLR instruments, upgrades will be required at the LLRO. Williams et al. (2020) performed simulations which showed that the LLR science results (e.g., beta and gamma of the parameterized post-Newtonian in general relativity and the Love Numbers) using the NGLRs will be several hundred times more accurate than those obtained with Apollo arrays as the LLRO operations and computer modeling improve. The aforementioned science results obtained using the Apollo arrays have already provided the best accuracy currently available. The current NGLR design will provide a return signal at the LLRO that is greater than the current return signal level at the Apollo 11 CCR. Moreover, the Apollo 11 retroreflector array, which has been sufficient for all of the current LLRO, is known to be dust-compromised which reduces the signal strength but has little impact on range measurements. In the future, we expect that there will be similar dust deposition on NGLR. However, the laser technology is improving far more rapidly than the anticipated dust accumulation. The accuracy of the Apollo science results depended critically upon the ability to obtain accurate range measurements over a long duration, i.e., many decades. It is essential that the NGLR be constructed on these demonstrated principles to ensure a long duration mission, well over 50 years.

Expanding the current laser ranging network on the Moon through the deployment of NGLR will yield new insights into the Moon, its properties, our understanding of gravitation, and of the accuracy of general relativity. Section 3.4 further describes expected NGLR science improvements that are up to hundreds of times more accurate than those obtained with existing retroreflector arrays. In many cases, the latter are already the most accurate results available.

### 2.3.3 Heat Flow

A heat flow probe makes two separate measurements of thermal gradient and thermal conductivity of the depth interval of the regolith penetrated. Heat flow is obtained as a product of these two measurements. The thermal gradient is obtained by measuring temperatures at multiple depths. The thermal conductivity is obtained in-situ by applying heat into the regolith and monitoring the rate of temperature increase (de Vries and Peck, 1958). In order to sense the flow of heat originating from the deep interior of the Moon, the probe must penetrate 2 to 3 m into the regolith, well below the reach of the thermal waves associated with annual and diurnal insolation cycles (Cohen et al., 2009).

The astronauts on Apollo 15 and 17 successfully obtained heat flow measurements by drilling holes into the regolith, using a rotary-percussive drill, and inserting thermal sensors (Langseth et al, 1976). However, each of these sites was close to the PKT boundary, so an unambiguous heat measurement of the PKT and the Feldspathic Highlands Terrane was not made. Deploying a heat flow probe semi-autonomously on a robotic lander mission such as the LGN would require a more compact system with less operational complexity. Such a system is currently under development (Nagihara et al., 2020). This system uses a pneumatic drill in penetrating into lunar regolith, which is more robust than the internal hammering technique employed for the heat flow probe on the InSight mission (Spohn et al, 2018). It integrates the thermal sensors into a pneumatic drill. As the probe penetrates into the regolith, it stops at multiple depths to make the thermal measurements on the way down to 3 m depth.

### 2.3.4 Electromagnetic

Electromagnetic (EM) sounding recovers the electrical-conductivity structure of the interior. Electrical conductivity is a strong function of temperature and, in the absence of free water, depends on composition-dependent ionic charge defects that for the Moon are most likely due to the abundance of ferric iron or hydrogen. EM sounding is complementary to measuring heat flow, in that the latter provides a boundary condition on the temperature profile derived from the former.

EM sounding was successfully performed during Apollo using the magnetic transfer function (TF), which compares the magnetic field at the target (the sum of source and induced fields) to a known, distant source field (see Sonett, 1982, for a review). These two quantities were measured by magnetometers at the Apollo 12 lunar surface site and the orbiting Explorer 35 satellite, respectively. Different formulations were used depending on whether the Moon was in the solar wind (Sonett et al., 1972), Earth's magnetotail (Hood and Schubert, 1978), or the lunar wake (Dyal et al., 1972). There was variation in the inferred temperature-depth curves: The first approach was consistent with thermal models for the Moon showing contemporary lower-mantle subsolidus convection (Hood and Sonett, 1982, Khan et al., 2006b). However, higher conductivities obtained in the last approach could require substantial internal melting. The Apollo 12 site was in the middle of PKT, calling into question whether its electrical-conductivity profile is representative of the bulk Moon (Grimm, 2013). A distribution of LGN magnetometers across PKT and FHT together with an orbital reference measurement would resolve differences in the principal terranes as well as providing global properties.

LGN can, however, exploit an alternative method that will yield more robust and better-resolved results, without requiring a reference orbiter. In the magnetotelluric method (MT), electrical-conductivity structure is determined by simultaneous measurement of the electric and magnetic fields at a single site (e.g., Simpson and Bahr, 2005; Chave and Jones, 2012). By eliminating the orbital measurement, data can be interpreted consistently across all three environments described above. Furthermore, MT is largely insensitive to finite wavelengths in the plasma that introduce distortions in the TF at frequencies of 10s of mHz and higher (Grimm and Delory, 2012). A larger bandwidth enables resolution of shallower structure than previous, particularly if the traditional fluxgate magnetometer is complemented with a search coil magnetometer. The system of surface electrodes used to measure the electric field is similar to those used in space physics (e.g., Bonnell et al., 2008) and naturally has broadband performance. A Lunar Magnetotelluric Sounder (LMS) has been selected for flight under NASA's CLPS program. In order to determine the electrical conductivity to as shallow as 200 km with 20% accuracy, LMS must measure electric and magnetic fields to 10 Hz with resolutions of 6 $\mu$V/m and 90 pT, respectively.

## 3. LANDING SITE SCIENCE RATIONALE

The general and measurement specific criteria for the LGN landing sites are outlined in Table 1. Any landing sites for the LGN mission should greatly expand the footprint of the prior Apollo geophysical network, including the farside, and use the knowledge gained by Apollo about the Moon to strategically investigate the nature of the lunar interior. A wider geographical spread of stations permits improved global structure determination from seismology and laser ranging. Including a station near the lunar limb, the edge of the visible disc of the lunar nearside hemisphere, maximizes the geographical extent between stations. The farside station will optimize new assessments of global seismicity and the structure of the Moon's lower mantle and core (see Section 3.3.2). The stations should maximize the recording of seismic signals from known deep moonquake clusters as they pass through the lowermost mantle and core (Figure 3, 4, and Yamada et al., 2013), and they should expand the current lunar laser retroreflector network (Figure 1). They should also be well within terrane boundaries to unambiguously make magnetotelluric and heat flow measurements both inside and well outside the boundaries of the PKT. Secondarily, landing sites should include: 1) be in the proximity of recently recognized thrust faults (lobate scarps) that could still be active and represent sources of shallow moonquakes (e.g., Kumar et al., 2016, 2019; Watters et al., 2019); 2) avoid local crustal magnetic anomalies that would contaminate the magnetotelluric sounding measurements; and 3) be on surfaces of sufficient age such that the regolith is suitably developed to allow the full deployment of the heat flow probe. Regolith thicknesses vary widely with generally thicker regolith on the farside and thicker regolith in the highlands versus the maria (e.g., Bart et al., 2011). Assessing an average rate of regolith formation will yield a rough order of magnitude estimate at best but is needed for planning purposes. Regolith formation and overturn are dependent on crater flux, the influence of secondary impacts, and the size of the impact (Costello et al., 2018). Various authors estimated regolith formation on the order of ~3–5 mm of new regolith per million years at ~3.8 Ga and about 1 mm/Myr from ~3.5 Ga to the present day (Oberbeck and Quaide, 1968; Quaide and Oberbeck, 1975; Hörz et al., 1991; Fa et al., 2014; Hirabayashi et al., 2018; Costello et al., 2018, 2020; Yue et al., 2019). This equates to 1 m of regolith formation per billion years. In order to facilitate heat flow probe deployment up to 3 m, a landing site on a surface ≥ 3 Ga is needed.

**Table 1**. LGN landing site requirements.

| | Landing Site Requirements | |
|---|---|---|
| **Overarching Criteria** | Broader coverage than Apollo | |
| | Low-risk landing sites (i.e., low boulder and crater density) | |
| **Measurement Type** | **Primary** | **Secondary** |
| Seismicity | Use known seismic sources to examine the core, mantle, and crust | Proximty to thrust faults (lobate scarps, wrinkle ridges) |
| Laser Retroreflectometry | 3 nearside stations | Place nearside stations near as possible to the limb |
| Magnetotelluric | Location well within terrane boundaries | Benign magnetic signature at the surface |
| Heat Flow | Location well within terrane boundaries | Regolith available for full deployment |

### 3.1 Landing Site Examples

For purposes of this paper, we have applied the landing site criteria (Table 1) to achieving the goal of the LGN mission. Therefore, example landing sites have been chosen to reflect a global distribution around the Moon, greatly increasing the footprint of the former Apollo network (Figure 1). Four landers are baselined, three on the nearside and one on the farside, with the sites chosen to fulfill the landing site criteria described above (Table 1). Rationale for these example sites are as follows:

*Procellarum KREEP Terrane* (P-5 basalt field, Hiesinger et al., 2003) site southeast of the Aristarchus Plateau (latitude = 14.9˚; longitude = -35.5˚): Relatively flat volcanic terrain, with few craters and boulders that has a crater size frequency distribution age of 3.48 Ga (Hiesinger et al., 2011a). This landing site is well within the boundaries of the PKT and close to the Th high just west of Copernicus crater (Figure 1). It is situated on basalts of sufficient age and with sufficient available regolith for full deployment of the heat flow probe. This example site is well-situated to detect both direct and core-reflected arrivals from the known nearside deep moonquake clusters (Figure 4). Lobate scarps are in this region to the east and southeast of this site (Figure 5).

*Schickard Basin* (latitude = -44.3˚; longitude = -55.1˚): This site is in the southern hemisphere of the Moon and the floor is partially flooded with basaltic lava flows (3.62-3.75 Ga; Hiesinger et al., 2011a) that form a relatively flat landing site, with few craters and boulder fields, but are old enough to have regolith thicknesses that allow full deployment of the heat flow probe. This example site is well outside the PKT, situated ideally to detect seismic phases reflected by the lunar core by known deep moonquake clusters in the north-eastern quadrant of the nearside, and refracted waves from the farside A-33 deep moonquake cluster through the core (Figure 3). It does not contain any local magnetic anomalies. Lobate scarps are in the region around the Schickard Basin (Figure 5). This example site also expands the LLR network into the southern hemisphere of the Moon.

*Crisium Basin* (latitude = 18.5˚; longitude = 61.8˚): The basaltic lavas on the floor of the basin form a relatively flat terrain, but contain secondary crater populations that will need to be avoided. According to the latest crustal thickness maps, the primary crust is essentially absent (Wieczorek et al., 2013) allowing mantle heat flow to be directly measured. Seismic measurements of the mantle at this site will also benefit from not being distorted by the fractured crust. For magnetotelluric measurements, the known magnetic anomalies within Mare Crisium (e.g., Richmond and Hood, 2008) need to be avoided. Therefore, this site is in the northeast section of the basin where the age of the basalts (based on crater size frequency distributions) are 3.2 Ga, although a 2.5 Ga unit is observed in the extreme northeastern sector of Mare Crisium (Boyce and Johnson, 1977; Hiesinger et al., 2011b). Note that Luna 24 returned basalts that yielded an age of ~3.65 Ga based on Rb-Sr systematics (Nyquist et al., 1978). This site is located close to wrinkle ridges within Mare Crisium and lobate scarps are located in the south of the basin and just to the east (Figure 5). This example site also expands the LLR network further to the east.

*Korolev Basin* (latitude = -2.4˚; longitude = -159.3˚): This site will allow the first surface geophysical measurements to be made on the farside of the Moon. The Korolev Basin is a Nectarian-age basin (3.85-3.92 Ga) and affords a relatively flat and boulder-free landing area that is in the vicinity of a lobate scarp. It is situated well within the Feldspathic Highlands Terrane and in the highest topographic area of the Moon, which represents the thickest crust (*cf.* Wieczorek et al., 2013). The site is approximately antipodal to many nearside deep moonquake clusters, again improving ray coverage for core-traversing seismic phases (Figure 4 and Yamada et al., 2013). This site will contain sufficient regolith for heat flow probe deployment and does not contain any local magnetic anomalies. A lobate scarp is close to the western margin of the basin (Figure 5).

## 3.2 Landing Site Descope Options

While the *baseline* LGN mission is for 4 identical landers to be distributed around the Moon, the *threshold* mission is for two landers: one at the PKT (P-5) site and one in the Schickard basin (Figure 1). This preserves deployment in distinct terranes with distinct crustal thicknesses and thermal regimes, uses known deep moonquake clusters (including those on the farside) to explore the deep lunar structure, and expands the current LLR network. Between the baseline and threshold missions a graceful descope trade space exists. The lander descope scenario presented here places the farside site as a high priority as this maximally enables observation of core-transmitted phases from the known nearside deep moonquake clusters (Yamada et al., 2013).

- **Descope** the Crisium lander. **Rationale**: Minimizes impacts to observations of seismic coverage for core-transmitted phases. Crisium Basin will have some geophysical data from the 2023 CLPS mission, although no seismic data will be returned. Removal of this site also removes expansion of the LLR network to the eastern part of the Moon.
- **Descope** two nearside landers (Schickard and Crisium). **Rationale**: Minimizes impacts to observations of seismic coverage for core-transmitted phases, and preserves deployment of landers in the PKT and the farside FHT. The remaining two sites permit using deep moonquake activity to explore the core and mantle of the Moon (the Korolev Basin site is antipodal to the nearside A-1 deep moonquake nest, and the PKT site is offset to the A-33 farside nest to explore mantle structure).
- **Descope** the communications satellite and deploy four nearside landers. **Rationale**: Reduces operational complexity, but has direct to Earth as the only communications option. Place the original farside lander at a polar region (e.g., in the Wiechert region of the south pole close to the observed lobate scarps). This removes any investigation of the thick crust on the farside, but retains the investigation of lobate scarps, heat flow in the FHT, and adds a NGLR station at the south pole.
• **Descope** the orbiter and farside lander, deploy three nearside landers in the sites proposed. **Rationale**: Reduces operational complexity, but has direct to Earth as the only communications option.
• **Descope** the Crisium and Korolev landers and the communications satellite. **Rationale**: this represents the threshold LGN mission. This preserves the ability to understand the internal structure of the Moon (utilizing deep moonquake cluster activity), records heat flow within the PKT and FHT, adds a southern hemisphere node to the NGLR network, and thus achieves the LGN mission goal.

### 3.3 Seismic network requirements

#### 3.3.1   Seismic ray density

The geometry of the seismic array is the predominant driver for LGN site selections. The LGN stations are situated to vastly improve our knowledge about the lunar deep mantle and core. Compared to Apollo, the wider aperture of the LGN array and geographical distribution of the stations provides ray path sampling of the entirety of the lunar interior. Using the seismicity catalog of Apollo, we calculated ray path densities for P (direct compression wave), PcP (core-reflected P-wave), and PKP (core-traversing P-wave) to the LGN across epicentral distances using the Weber et al. (2011) velocity model and proposed LGN station locations (Figure 3). Consistent seismic ray path density across epicentral distance is crucial for providing a full picture of lunar internal structure. The LGN provides a significantly denser sampling for core traversing waves (e.g., PKP), waves that sample the deep mantle (PcP, P), and more uniform sampling of the crust and upper mantle than Apollo. The more complete coverage is enabled by the deployment of four stations, particularly from a farside station in Korolev crater that provides deep mantle and core sampling and will allow for the detection of farside seismicity that was unobserved by Apollo.

Ray path density through depth in the lunar interior is obtained by taking event hypocenters from Gagnepain-Beyneix et al. (2006) and raytracing to the locations of both the LGN stations (Figure 1) and the Apollo stations for comparison. Ray paths are calculated assuming the Weber at al. (2011) velocity model using the TauP Toolkit (Crotwell et al., 1999) for all sources and the phases P, PcP (core reflections), and PKP (core traversals). Each event/station pair is aligned on the great circle path and arranged by epicentral distance from individual stations to illustrate sampling density with depth (not accounting for 3D sampling geometry).. Apollo stations primarily sampled the uppermost mantle and crust of the Moon. By comparison, the LGN station geometry, assuming similar seismicity, would provide at least double the increased sampling density within the core and lowermost mantle, and extend sampling to rays traversing to greater epicentral distance.

To quantify LGN's increased coverage at a given distance, consider the observed distribution of events of the Apollo event catalog. Ray paths that are sensitive to PKP range from 180-270°. For PKP, LGN will observe 100% of events at these distances with at least 1 event per 5°, whereas the Apollo PSE only had 55%. LGN covers 89% of these distance with 2 or more events per 5° and 72% for 3 or more events per 5°. In contrast, Apollo covered only 16% for 2 or more events and 11% (using one small distance window of ±5°) for 3 or more. LGN expands the distance range covered by the Apollo network to a truly global sampling of ray paths. This is a conservative estimate assuming that LGN were only to record the distribution of seismicity seen by Apollo, however, extrapolating to a larger collection area and improved instrument sensitivity will improve ray coverage statistics. The long term periodicity of the deep moonquake clusters has been determined to still be active today (Weber et al., 2010).

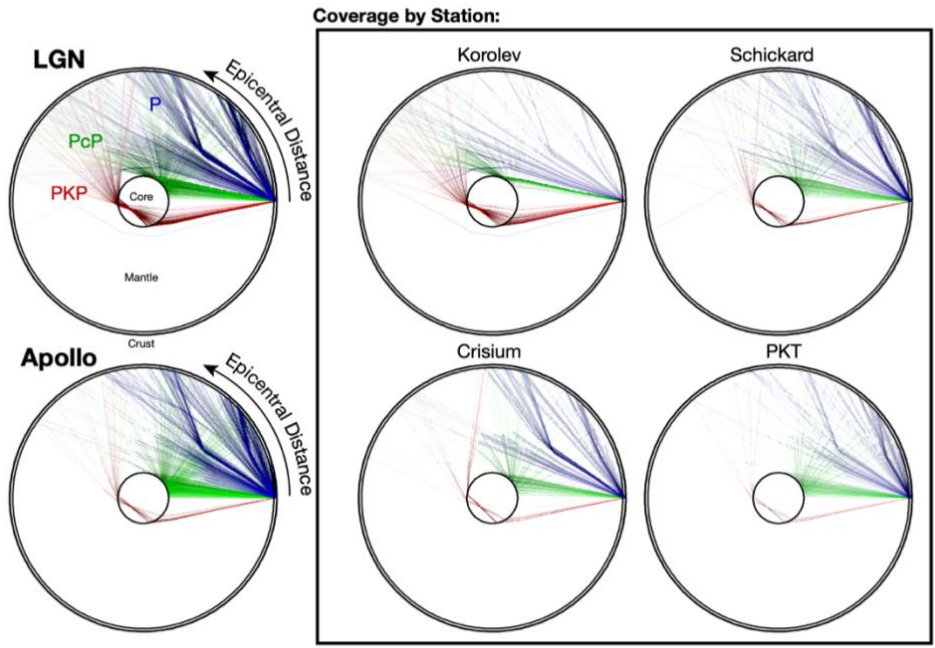

**Figure 3**. Ray path coverage for P (blue), PcP (green), and PKP (red) phases from the known distribution of deep moonquake clusters, rotated such that the station is fixed at 0 degrees and stacking all stations in the array for LGN (top left) and Apollo (bottom left). Note LGN's wider coverage in epicentral distance, and denser sampling of the deep interior. The box on the right shows ray coverage for each of the LGN stations separately.

### 3.3.2    Location Optimization for deep structure determination

Previous work suggests that networks with large distances between stations provide the best configurations for deep structure determination, including the core (Yamada et al., 2011). Accordingly, the mean distance between the stations of the notional LGN network is therefore around 105°, including a station on the lunar farside. The farside station serves two additional purposes: first, it permits detection of farside seismicity, which was seldom observed by the nearside Apollo array (Nakamura, 2005); and second, it increases detectability of secondary seismic phases from the known distribution of deep moonquake clusters that traverse the core (Figure 4). To discriminate between these core phases and the direct P-wave arrival, the farside station should be located far from the lunar limbs. The Korolev station location is consistent with these requirements.

The nearside stations are closer to the limbs, which will permit the detection of direct P waves from meteoroid impacts. These occur preferentially closer to the leading limb (Le Feuvre & Wieczorek., 2011), and are observed at both limbs by the ground-based impact flash monitoring program (Moser et al., 2015). These impact flash detections provide a reference timing and location for the events recorded on the seismometers, and will provide strong constraints on estimates of crustal structure around these stations.

Core phase detection is illustrated in Figure 4 (adapted from Yamada et al., 2013), which shows the number of ScS (core-reflected shear wave) and PKP phases detected per year at 0.1 Hz by a VBB instrument with noise specifications indicated in Section 2.3.1 (3.5e-11 m/s/s/√Hz at 0.1 Hz), from 15 of the most active deep moonquake nests. This computation takes into account the number of events per year at each nest and the distribution of their moment magnitudes as measured by the Apollo network (see Yamada et al., 2013), and assumes the Very Preliminary REference MOON model (VPREMOON) seismic velocity and attenuation model (Garcia et al., 2011). The maximum ground velocity of each phase is computed at 0.1 Hz and a detection is counted when this value is five times larger than the root mean square instrument noise in the 0.07 Hz to 0.14 Hz range, which is estimated to be 1.49e-11 m/s. Due to the refocusing of seismic waves at the antipode of their source, seismic events with the lowest seismic moment allow PKP detections only at their antipode, whereas larger seismic moments allow detections over a large epicentral distance range. These numbers represent a lower bound because additional detections may be performed from quakes already located by Apollo but observed with a higher Signal to Noise Ratio (SNR) by LGN, as well as newly detected quakes located only by LGN stations.

The stations on the nearside are located to adequately detect ScS phases from already known deep moonquake locations. The farside Korolev crater location is situated among the antipodes of a high number of nearside deep moonquakes, thus optimizing the number of PKP phase detections over a large epicentral distance range. The uncertainties on the deep moonquake moment magnitude translate into an error bar on the optimized locations of seismometers of at least 5 degrees. The LGN sites are positioned to capture the required minimum of at least one path for core traversing PKP seismic waves generated by an Apollo-detected DMQ clusters, and also provides a chance to see PKP over a range of distances that will enable determination of core velocity structure and layering. Note that the farside Korolev station, in the event that it detects farside seismicity, will also be able to record core-reflected phases from that seismicity.

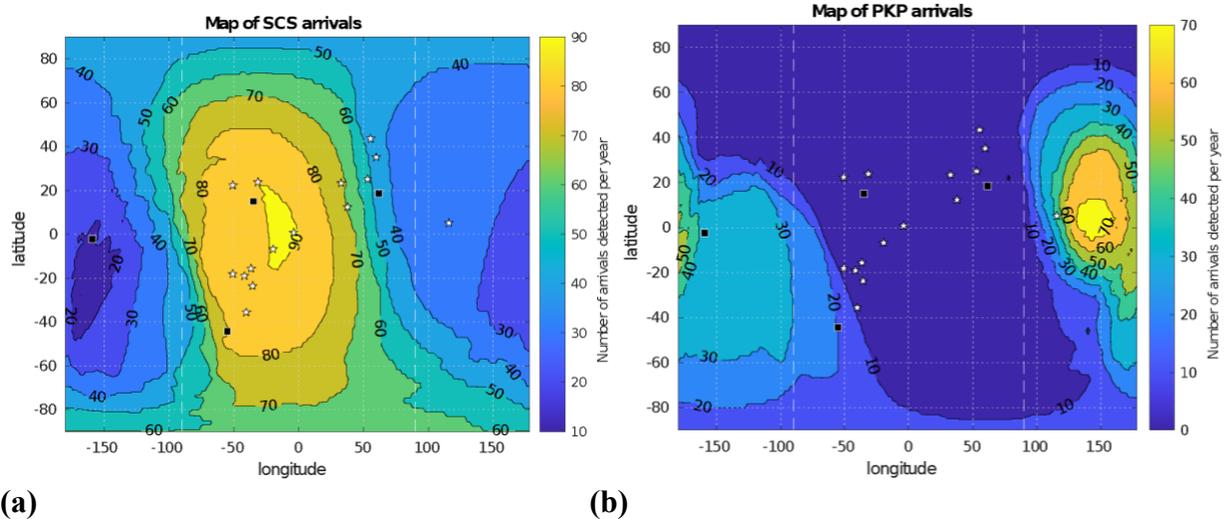

**(a)**                                          **(b)**

**Figure 4.** (a) Number of core-reflected shear-wave (ScS) phases and (b) Number of core-traversing P-wave (PKP) phases detected per year (color bar) from 15 energetic deep moonquake nest epicenters (white stars) as a function of the location of a VBB seismometer. Black squares indicate LGN station locations. The number of (ScS; PKP) core phases detected at these positions are (87; 2), (78; 20), (45; 0) and (17; 35) for PKT, Schickard, Crisium and Korolev stations, respectively. Vertical dashed lines mark the limit between farside and nearside. Computed for a RMS noise level of 1.49e-11 m/s in the 0.07-0.14 Hz frequency range following the method described in Yamada et al. (2013).

### *3.3.3 Lobate scarps*

It has long been held that lunar tectonics, the deformation of the Moon's near-surface crustal materials, was largely restricted to the mare basins, and that this deformation was ancient. This view has been radically altered by high resolution Lunar Reconnaissance Orbiter Camera (LROC) images collected over the last decade. A globally distributed population of lobate scarps has been revealed in these images (Figure 5) (Watters et al., 2010, 2015, 2019). Lunar lobate scarps are small-scale landforms that are the surface expression of thrust faults, contractional faults that displace crustal materials up and over adjacent terrains (Binder, 1982; Watters and Johnson, 2010). Even more remarkable is the age of these fault scarps – their small size, pristine appearance, lack of superimposed impact craters, and crosscutting relations with small-diameter craters – all suggesting they are very young (Watters et al., 2010). The size-frequency distributions of impact craters proximal to the scarps and erasure of crater populations <~20-100 m in diameter up to kilometers away shows the fault scarps were actively forming in the late Copernican (<400 Ma) (van der Bogert et al., 2018). A very young age is also estimated from infilling rates for small-scale back-scarp graben, suggesting the fault scarps were active within the last 50 Ma (Watters et al., 2012; Kumar et al., 2019).

To date, over 3,500 lobate scarps have been identified on the Moon (Watters et al., 2019) (Figure 5). Although they occur in both the highlands and mare terrains, they are most commonly found in the highlands where they are the dominant tectonic landform. The large number of fault scarps, their global spatial distribution, and their very young age, has led to the hypothesis that the faults are currently active and could be the source of some moonquakes recorded by the Apollo Seismic Network. Twenty-eight shallow moonquakes (SMQs), with hypocenter depths generally constrained to be <100 km, are among the strongest recorded events (Nakamura et al., 1979; Khan et al., 2000; Gillet et al., 2016). Contrary to this previous work, Watters et al. (2019) postulate a much shallower origin for SMQs (< 1 km). A comparison of relocated candidate epicenters with surface solutions only, generated by an algorithm adapted for sparse seismic networks, indicates that 8 out of 17 SMQs with surface solutions fall within 30 km of a mapped lobate scarp (10 inside of 60 km), the closest within 4 km (Watters et al., 2019). Shake models predict moderate to strong ground shaking over a distance of 30 to 60 km (Watters et al., 2019). In addition, 7 SMQs within 60 km of a lobate scarp occur near the regions where stress models predict peak compressional stresses (Watters et al., 2019). The proximity and timing of SMQs supports the hypothesis that they are due to slip events on lobate scarp-related active thrust faults.

Additional criteria for selecting the locations of the LGN landing sites is their proximity to potentially active faults (see Table 1). The nominal PKT site is ~460 km from one of the most prominent lobate scarps found in mare basalts. The Herodotus scarp (informal name) occurs near the southern margin of the Aristarchus Plateau and has a maximum relief of ~100 m. The site is also ~110 km from a wrinkle ridge-lobate scarp transition in Procellarum. These transitions occur at the contact between mare basalt and highlands or highland massifs and reflect the response of the thrust fault to the contrast in mechanical properties (Watters and Johnson, 2010). The highlands within ~500 km of the Schickard basin landing site have a large number of lobate scarps and scarp clusters. This includes the Vitello cluster, one of the longest clusters of lobate scarps found on the Moon. The closest lobate scarp is on the western rim of Schickard, ~90 km from the nominal landing site. The Mare Crisium landing site is dominated by wrinkle ridges. However, on the southern margin of Crisium, a wrinkle ridge-lobate scarp transition is found ~250 km from the proposed landing site. The Korolev basin landing site is near another large cluster of lobate scarps. Outside the northwest rim, ~180 km from the nominal landing site, is the Korolev cluster. This prominent cluster of scarps covers over 140 km of terrain. Thus, the proposed LGN landing sites are well placed to detect coseismic slip events on lobate scarp thrust faults.

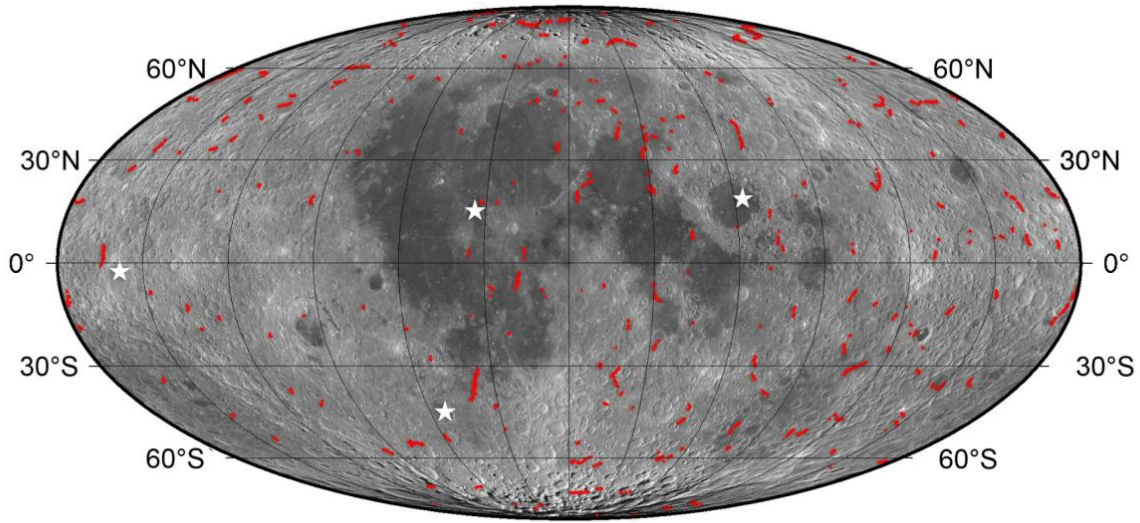

**Figure 5.** Global distribution of lunar lobate scarps. These lobate thrust fault scarps (red lines) are found predominantly in the highlands, at all latitudes. Over 3,500 of these young fault scarps have been detected thus far in Lunar Reconnaissance Orbiter Camera (LROC) images. LGN proposed stations are marked (stars).

### 3.4 Geodetic station optimization

To understand the expected science return of a NGLR network, we performed simulations with varied number of stations and LLRO measurement precision. These results are compared to the Apollo CCR array for a variety of science parameters and factors of accuracy improvement are tabulated in Table 2. Six years of synthetic LLR data from the network of proposed nearside sites are considered. We assume the use of the current range accuracies for the Apollo retroreflector arrays. The accuracy of a normal point, which requires multiple ranges, obtained by the LLRO depends on the hardware and observing procedures at that location.

The first row considers the improvement in the accuracy with respect to the LLR accuracy obtained with the current retroreflectors under the assumptions of the three NGLR and LLRO normal point accuracies of 1.5 mm. Precision at this level has already been obtained at the Apache Point Observatory Lunar Laser-ranging Operation (APOLLO) LLRO. The second row addresses the improvement in the factors when a fourth NGLR is deployed at the south pole (i.e., the Artemis III site). A second set of simulations (Williams et al 2020) is shown in the third and fourth rows for 3 and 4 NGLRs with range accuracies of 0.15 mm, addressing the limiting accuracy that the NGLR will support in the current design of the LGN deployment. The basic NGLR package can support very high accuracy ranging (Currie et al. 2013). The accuracy improvement for a given science result is estimated for a single range for the basic NGLR package (Currie et al. 2013), with the normal point being obtained by multiple ranges. These results require an upgrade in LLRO hardware. In order to reach the goal of 0.15 mm accuracy, the horizontal gradients of temperature, density and humidity in the Earth's atmosphere must be determined either by modeling or by direct measurement. The modeling approach has been demonstrated using data from a variety of satellite laser ranging (SLR) stations (Drożdżewski et al 2019; Masoumi et al 2017) although it has not yet been implemented at any of the LLRO.

Improved ranging accuracy is achieved by expanding the footprint of LGN's NGLR and translates into improved science results as a function of the number of NGLR deployed (3 or 4 stations) and status of the LLRO. In particular, the landing sites far from the equatorial sub-Earth point, like the example LGN landing sites or a nearly polar site such as the nominal South Pole site of the Artemis III mission (Longitude = 0, Latitude = -88), are considered. The simulations assumed that the NGLR are deployed at Mare Crisium (Longitude: 59, Latitude: 17), a Western Site region (Longitude: -50, Latitude: 20), a south western site (Longitude: -55, Latitude: -45), which are the sites expected to the explored by missions of the NASA CLPS program with NGLR funded by the NASA Lunar Surface Instrument and Technology Payloads (LSITP) program, and the south pole Artemis III site. We expect that the LLR science improvements due to any of the nearside LGN sites to be comparable. If both LGN and CLPS deployments are successful, along with the currently existing retroreflector arrays the science results will be further improved, particularly the Love Numbers and the selenodetic coordinate system. The latter will improve the accuracy of maps of various properties of the lunar surface, particularly catalogs like that from LRO. Deployments near the limbs and/or the poles, as compared to the current sites, will improve science accuracy by a factor of 3 to 5.

The science parameters listed in Table 2 include: moment combinations b=(C−A)/B and g=(B−A)/C, the distortions of the lunar crust in response to the tidal forces (Love Numbers $h_2$ and $l_2$), a parameter (cos D) that addresses the general relativistic weak equivalence principle, the total mass of the Earth-Moon system that depends on the mean semimajor axis (<a>), and the longitude librations (Tau) at 815 days. All values in Table 2 are positive indicating improvement over previous. The simulations show that similar improvements by factors of hundreds in the science results with respect to the Apollo results can be obtained with the LGN NGLR deployments. This demonstrates the expected high degree of improvement in the accuracy of these science parameters with the LGN NGLR.

The Apollo results are based on ranges obtained from the past and current LLRO (see https://ilrs.gsfc.nasa.gov/network/stations/index.html for global map of stations) ranging to the existing retroreflector arrays (or have contributed significant ranging data in the past). These consist of LLRO located in the US, i.e., McDonald (MCD) (Silverberg & Currie, 1971) and (MLRS), Hawaii (Hall) and Apache Point (APOLLO) (Murphy et al. 2004), France (MeO) (Chabé et al 2020), Italy (Matera) (Varghese et al 1993), Germany (Wettzell).

**Table 2.** NGLR simulated accuracy improvement over current for 3 and 4 nearside locations, seven science parameters, and at 1.5 or 0.15 mm LLRO normal points precision. The 3 nearside locations are the three 3 CLPS landing sites. 4 sites include these and the Artemis III site. The science parameters encompass geophysics and astrophysics analyses (as described in text). The simulations assume the LLRO currently performing ranging (APOLLO, MeO, Matera and Wettzell), and their rates of ranging operations. The current NGLR design will support the best accuracy (0.15 mm) showing consistent increase in science return.

| # of sites | Accuracy | Beta | Gam | h₂ | l₂ | cos D | <a> | Tau |
|------------|----------|------|-----|-----|-----|-------|-----|-----|
| 3 | 1.5 mm | 4 | 7.2 | 5 | 5 | 2.8 | 3.2 | 3.5 |
| 4 | 1.5 mm | 4.4 | 7.4 | 6.8 | 13 | 3.2 | 4.2 | 3.6 |
| 3 | 0.15 mm | 100 | 410 | 20 | 27 | 145 | 120 | 31 |
| 4 | 0.15 mm | 111 | 420 | 212 | 570 | 162 | 147 | 330 |

### 3.5 Heat Flow observation considerations

In understanding the Moon's thermal evolution, it is important to quantify the contribution of crustal radiogenic heat to the overall heat budget of the Moon and its geographic variation. The Apollo program considered heat flow measurements a high priority and originally planned to carry out measurements by astronauts on four of the landed missions. However, only two of them (Apollo 15 and 17) made successful heat flow measurements (Langseth et al, 1976). Since then, a number of expert panel studies have recommended additional heat flow measurements for future lunar-landing missions (National Research Council, 2007; Cohen et al., 2009; National Research Council, 2011; Lunar Exploration Analysis Group, 2017), because the two existing measurements alone are not sufficient.

The heat flow values obtained by the Apollo 15 and 17 astronauts were 21 mW/m$^2$ and 16 mW/m$^2$, respectively (Figure 6). It has been hypothesized that the higher value at the Apollo 15 site was due to its location being within the PKT with higher concentrations of radionuclides in the crust (Wieczorek and Phillips, 2000). To further test this hypothesis, we should examine whether or not surface heat flow values have a positive correlation with the surface abundance of radionuclides. Besides the surface abundance of radionuclides, other factors such as crustal thickness (Figure 7) and lithology may also be controlling the radiogenic heat production within the crust.

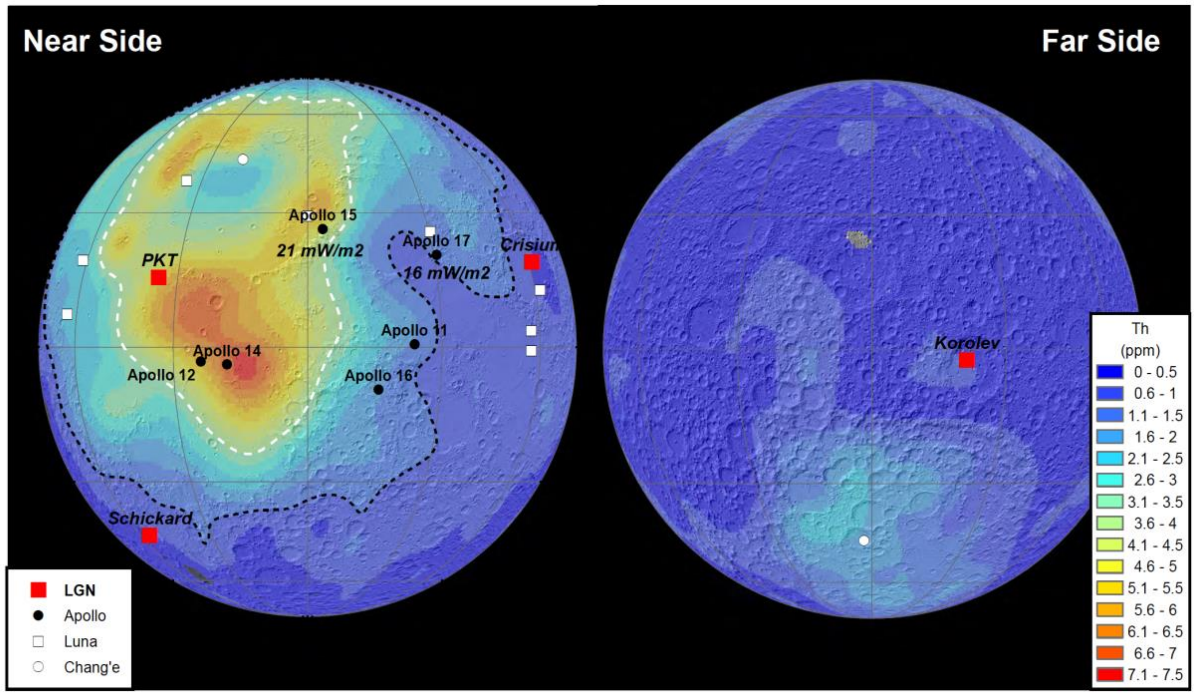

**Figure 6.** Near- and Farside maps of surface Thorium concentration (ppm) obtained by the Kaguya mission (https://darts.isas.jaxa.jp/planet/pdap/selene/index.html.en). The red squares indicate the proposed landing sites for the LGN mission. The landing sites for the Apollo, Luna and Chang'e missions also shown by circles and squares. The white dashed line is the 3.5 ppm contour and roughly delineates the geographic extent of the PKT. The black dashed line is the 1-ppm contour which can be used as the minimum geographic extent of the area where heat flow may be affected by Th-rich ejecta from the Imbrium basin-forming impact. The heat flow values at the Apollo 15 and 17 sites are also shown.

We should also acquire a data point in an area where crustal radiogenic heat is expected to be minimal in order to establish a baseline for the Moon. Such a baseline value should closely approximate the heat flow out of the lunar mantle. The mantle heat flow estimate, combined with the magnetotelluric sounding and the seismic investigation, would enable us to unambiguously constrain the thermal properties and composition of the deep interior of the Moon.

The four landing sites proposed for the LGN mission, together with Apollo 15 and 17, will allow us to sample a large range of radionuclide concentrations, crustal thickness and lithology (Figure S1). The 'PKT' site and Apollo 15 are located within areas of high Th concentration (4.5 – 5 ppm) with variable crustal thicknesses (25 and 39 km, respectively). 'Schickard', 'Crisium', and 'Korolev' are located far outside of the PKT with very low Th concentrations (0.5 ppm). Crisium has a very thin (< 10 km) mare crust, and a heat flow measurement there could establish the baseline for the Moon. Korelev and Schickard have highland crust with 58 and 42 km thick crust, respectively.

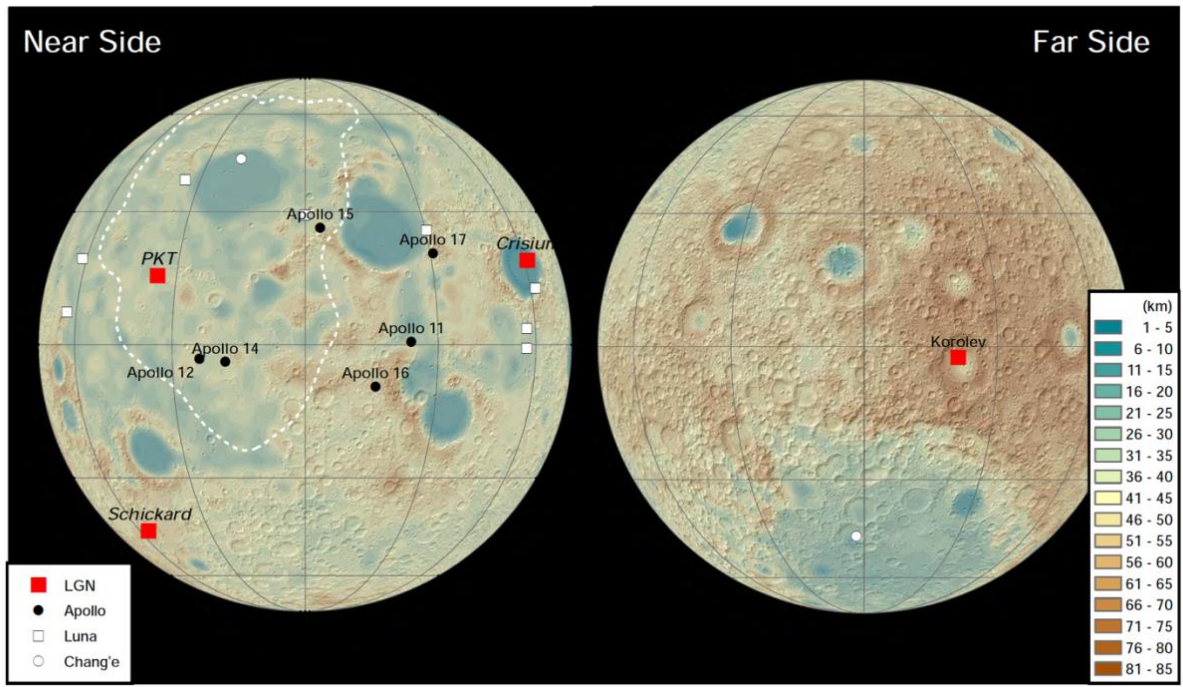

**Figure 7.** Near- and Farside maps based on the crustal thickness estimates (km) obtained by the GRAIL mission (Wieczorek et al., 2013). The red squares indicate the proposed landing sites for the LGN mission. The landing sites for the Apollo, Luna and Chang'e missions are also shown by circles and squares. The white dashed line is the 3.5 ppm contour and roughly delineates the geographic extent of the PKT.

### 3.6 Electromagnetic network requirements

Previous electromagnetic experiments, performed at Apollo landing sites, were within or adjacent to the PKT. Therefore, it is unknown whether or not the deep electrical conductivity profile derived at Apollo 12 (see Sonett, 1982, for a review) is representative of the whole Moon. Comparison of sites in both PKT and FHT can test specific hypotheses for the lunar interior described above. If the electrical conductivity is lower beneath the FHT, then additional heat sources indeed likely exist in the PKT upper mantle. If the sub-FHT conductivity is higher than PKT, then there is no anomalous mantle heating, and either the PKT mantle is depleted in Fe or $H_2O$ compared to FHT or the FHT mantle is actually hotter due to thicker, insulating crust (in spite of likely lower mantle heat flow).

Additional sites within PKT are also useful. Mare Imbrium is a particularly diagnostic site (Nagihara and Grimm, 2020), as its thin crust and low surficial thorium suggest that crustal KREEP was excavated by the basin-forming impact. If the heat flow is comparable to the Apollo 15 measurement, then a mantle heat source is confirmed and the crustal geochemical signature is only incidental to thermal evolution. Conversely, significantly lower heat flow would point to the dominance of crustal radioactivity. In both cases, the electrical conductivity profile can be anchored on the heat-flow measurement to discern and compare upper-mantle temperatures.

## 4.  MARE CRISIUM DETAILED SITE SURVEY - CASE STUDY

We illustrate the details of site selection within the Mare Crisium basin, with emphasis on the requirements primarily from heat flow and electromagnetic sounding. Mare Crisium is desirable because it is not only outside of PKT, but lies on a great circle from the middle of PKT through the Apollo 15 and 17 sites where heat flow was previously measured. Furthermore, Mare Crisium is underlain by some of the thinnest crust on the Moon, so the heat-flow signature there will be representative of the bulk Moon, which can be linked along a great circle back into the PKT. On the basis of these factors advocated by the Lunar Instrumentation for Subsurface Thermal Exploration with Rapidity (LISTER) and Lunar Magnetotelluric Sounder (LMS) teams, Mare Crisium has been selected for the CLPS 19D mission landing site (with other payloads that are site-agnostic). This site for use within the LGN network will be re-evaluated following a successful CLPS 19D mission in the 2023 time frame. Heat flow, Magnetotellurics, and NGLR would all benefit from observing different locations than previous sites.

Other criteria for the specific landing site include: (1) flat surface with low rock abundance, (2) thick regolith, and (3) small static magnetic field. Criterion 1 is for landing safety and low probability of interference with deployments (both on the surface and in the subsurface). Criterion 2 is to assure that the heat-flow probe can penetrate up to several meters into the subsurface. Criterion 3 is to ensure that the measurements of time-varying magnetic fields can be performed in the highest sensitivity range and that the incoming fields are not locally distorted severely from plane-wave equivalents.

General Strategy. In selecting an exact location for landing the spacecraft in each of the aforementioned four areas proposed for the LGN mission (Figure 1), we consider science, engineering, logistical, and safety requirements for the individual payload instruments and the spacecraft. In identifying localities that satisfy these requirements, we use remote observations from either Earth or spacecraft orbiting around the Moon (Table 3).  Here we use Mare Crisium as a case study for how each LGN detailed landing site analysis will be discussed.

Crustal Thickness. Mare Crisium is chosen as one of the four landing areas for the LGN mission, because it is an ideal location for constraining the thermal structure of the mantle away from the PKT. Heat flow through the surface of the basin should be least affected by the radiogenic heat production within the thin (~10 km), mafic crust (Wieczorek et al., 2013; Arivazhagan and Karthi, 2018). In narrowing down the potential area for landing, we start with the area inside the 10 km crustal thickness contour (Figure 8a). This wide area of relatively flat topography and little variability in crustal thickness and composition should make geologic interpretation of the data acquired by the LGN mission relatively straightforward.

Magnetic Anomalies. To ensure that the measurements of time-varying magnetic fields by the magnetotelluric instrument can be performed in the highest sensitivity range with minimum distortion of the incoming time-varying fields by any localized plasma, the lander should avoid obvious sites of high static magnetic field and plasma-surface interaction, namely, swirls. Because surface fields at these locations could be hundreds to thousands of nT, we seek sites <100 nT as a conservative bound. Mare Crisium has 'bullseye' magnetic anomalies on both the north and south margins (Figure 8b). The maximum total field at 30 km altitude ~6 nT for the northern anomaly is modeled to be a maximum 60 nT at the surface (Baek et al., 2017; see also Tsunakawa et al., 2014).  However, earlier modeling of the northern source (Hood, 2011) as well as simple comparison of measured surface vs orbital magnetic fields (especially the Descartes swirl near Apollo 16: Dyal et al., 1973) suggests surface fields in Mare Crisium could be as high as several hundred nT. Nonetheless, suitable sites <1 nT at 30 km, and therefore likely <10-100 nT at the surface, can be identified in Figure 8b.

Regolith Properties: An optimal landing site must also have a relatively thick accumulation of regolith devoid of large rocks so that the heat flow probe can penetrate to 2 to 3 m depth. Presence of large rocks exposed on the surface can be inferred from the observation of the diurnal swing of thermal infrared waves, and that is how the rock abundance estimates (Figure 8c) are obtained from the Diviner instrument onboard NASA's LRO (Bandfield et al., 2011). To infer subsurface properties of the regolith, observation of the Earth-based, long-wavelength radar returns has been found effective. Radar waves can penetrate ~10 times the wavelength into the subsurface. Fa and Wieczorek (2012) estimate the regolith in Mare Crisium to be 4 to 7 m thick throughout, based on the P-band (70 cm wavelength) radar returns received at the Arecibo Observatory.

Buried blocky objects scatter and reflect radar signals with wavelength up to 10 times their size. Regolith relatively devoid of large rocks in the 1 to 7 m depth range show up as dark areas in circular polarization ratio (CPR) images of the S-band (12.6 cm wavelength) and P-band radar returns. It has been noted by previous observations (Ghent et al., 2010; Ghent et al., 2016) that fine-grained ejecta deposited some distance away from medium-to-large-size craters show up as 'dark haloes' in P-band CPR images. In Mare Crisium, such haloes are observed around Picard and Peirce (Figure 8d). Surface rock abundance is also low in these radar-dark haloes (Figure 8c).

Based on these observations, we have chosen two areas of 10 km radius for further consideration for landing the spacecraft. The one is located in the radar-dark halo of Picard, which is sufficiently far away from the two bullseye magnetic anomalies. The other is located in a radar-dark area not associated with craters in the central basin where crustal magnetic anomalies are the lowest (Figure 8a,c).

Surface Topography. Selecting a landing site within either of these two localities requires detailed knowledge of the topography. Safe landing of the spacecraft requires a wide, smooth, and level surface. Most of the payload instruments also require wide, flat, and level surface for their optimal performances; the seismometer for good mechanical coupling with the ground, the retroreflector for an unobstructed line of sight to Earth, and the heat flow probe for penetrating vertically into the regolith.

In identifying areas of smooth, level surface, we first utilize the digital elevation model (DEM) that combines data from the Lunar Orbiter Laser Altimeter (LOLA) onboard LRO and the laser altimeter onboard Kaguya (Barker et al., 2016). This DEM has global coverage with 60 m spatial resolution, and is well suited for identifying large-to-medium-size craters and other major positive and negative relief, basin related, tectonic features such as wrinkle ridges and graben that may be hazardous to landers (Figure 8e).

Even though much of the surface of the Mare Crisium appears smooth on these DEM, higher spatial resolution imagery (1-2 m, LROC Narrow Angle Camera, NAC) reveal that the surface is peppered with a number of small-diameter (< 10 m) craters. The LRO team is planning to acquire NAC stereo images over our two candidate areas in 2020 – 2021. For selecting an exact landing location, we will use the digital terrain model (DTM) derived from the stereo images.

**Table 3.** A list of the criteria and the remote sensing data used for landing sites consideration.

| Criteria | Instrument* | Available Data set | Spatial Resolution |
|---|---|---|---|
| Wide, smooth, and level surface | SC, S, HF, NGLR | SLDEM (LOLA and Kaguya) LROC-NAC DTM | 60 m ~2 m |
| Regolith thickness > 4 m | HF | Arecibo P-band radar | 400 m |
| Fine-grained regolith | HF | Arecibo P-band radar Arecibo S-band radar LRO-Diviner rock abundance | 400 m 80 m 240 m |
| Avoid large (<100 nT) magnetic anomalies | MT | Lunar Prospector, Kaguya-LMAG | 1° |
| Relatively uniform crustal thickness | S, HF, MT | GRAIL | 0.25° |

*SC: spacecraft, S: seismometer, HF: heat flow probe, NGLR: Next Gen Lunar Retroreflector, MT: magnetotelluric instrument

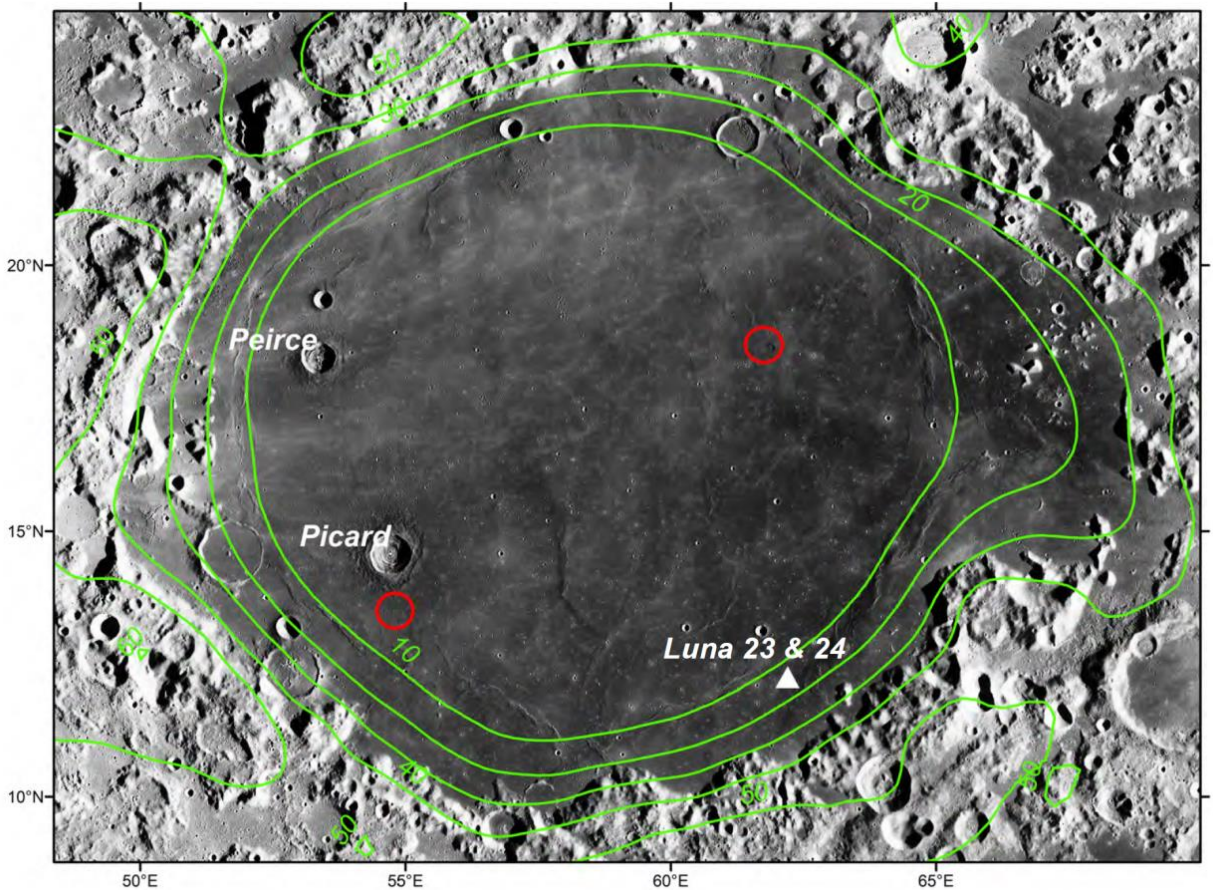

**Figure 8.** Contours (10 km intervals) of crustal thickness estimated from the GRAIL data (Wieczorek et al., 2013) drawn over mosaic of images obtained by the LRO Wide Angle Camera (WAC) of Mare Crisium. The red circles indicate the two candidate areas for further consideration for landing site selection. The white triangle indicates the landing sites of Luna 23 and 24. The two spacecrafts landed within 2-3 km of each other (Lawrence, 2013). Only Luna 24 was able to return samples to Earth.

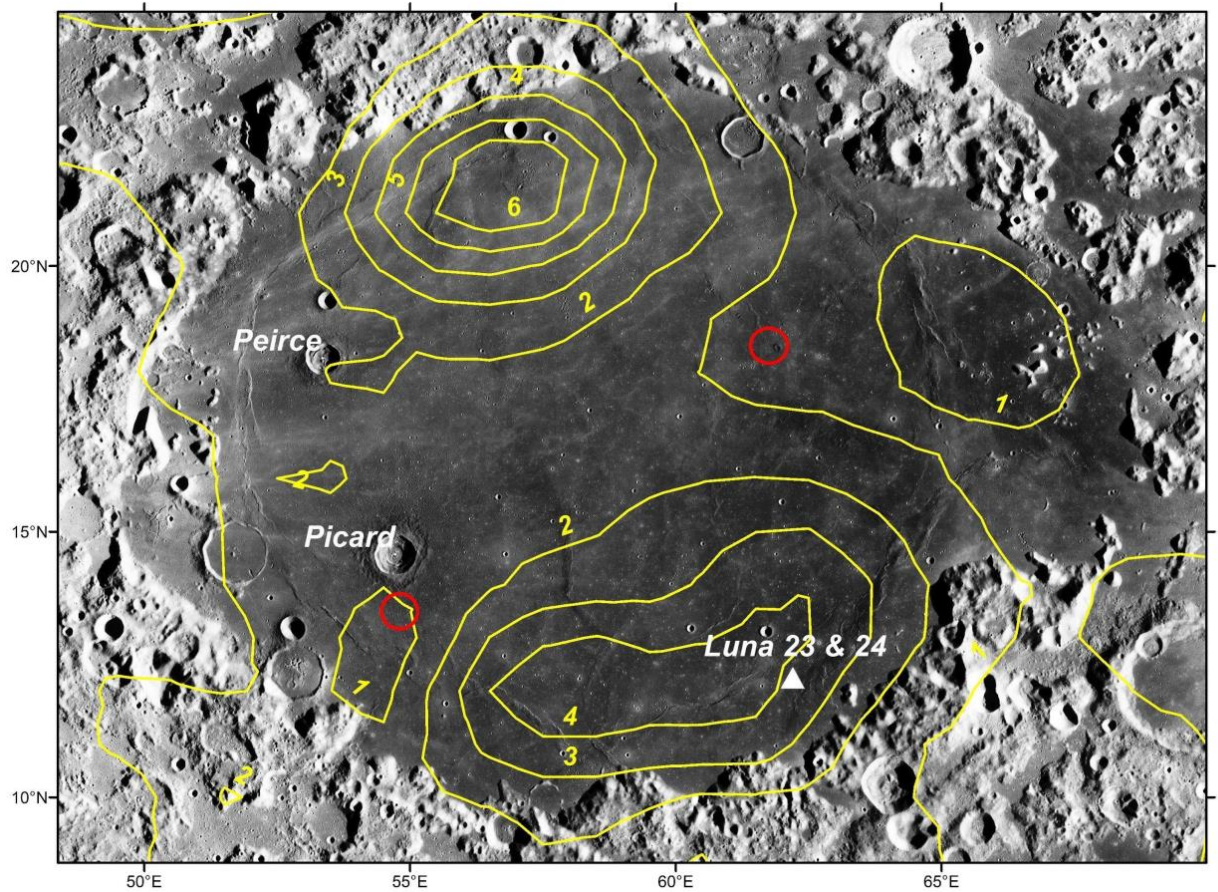

**Figure 9.** Contours (1 nT intervals) of magnetic anomaly intensity obtained at 30 km altitude by the Kaguya spacecraft (http://darts.isas.jaxa.jp/planet/pdap/selene/index.html.en), drawn over LRO WAC mosaic of Mare Crisium. The red circles indicate the two candidate areas for further consideration for landing site selection. The white triangle indicates the landing sites of Luna 23 and 24. See text for description of surface fields.

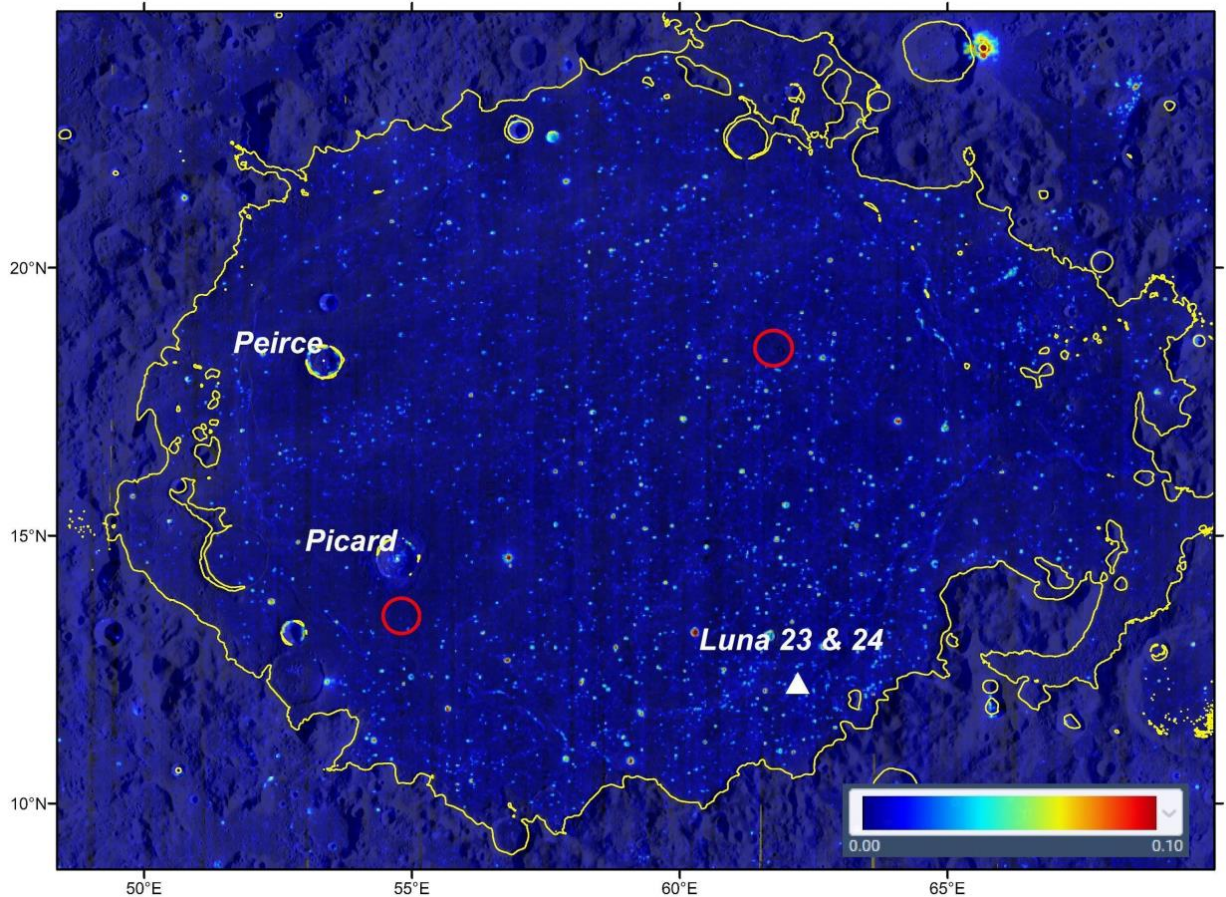

**Figure 10.** A map of rock abundance or concentration estimates derived from observations of the diurnal temperature swings by the LRO-Diviner instrument (Banfield et al., 2011). For each pixel, the fraction of the rock-covered area is given (0 to 0.1 color bar). Yellow lines are -3 km elevation contours that roughly define the extent of mare basalt fill within Crisium. The red circles indicate the two candidate areas for further consideration for landing site selection. The white triangle indicates the landing sites of Luna 23 and 24. Map data from Lunar QuickMap, available at https://bit.ly/32KYV5O.

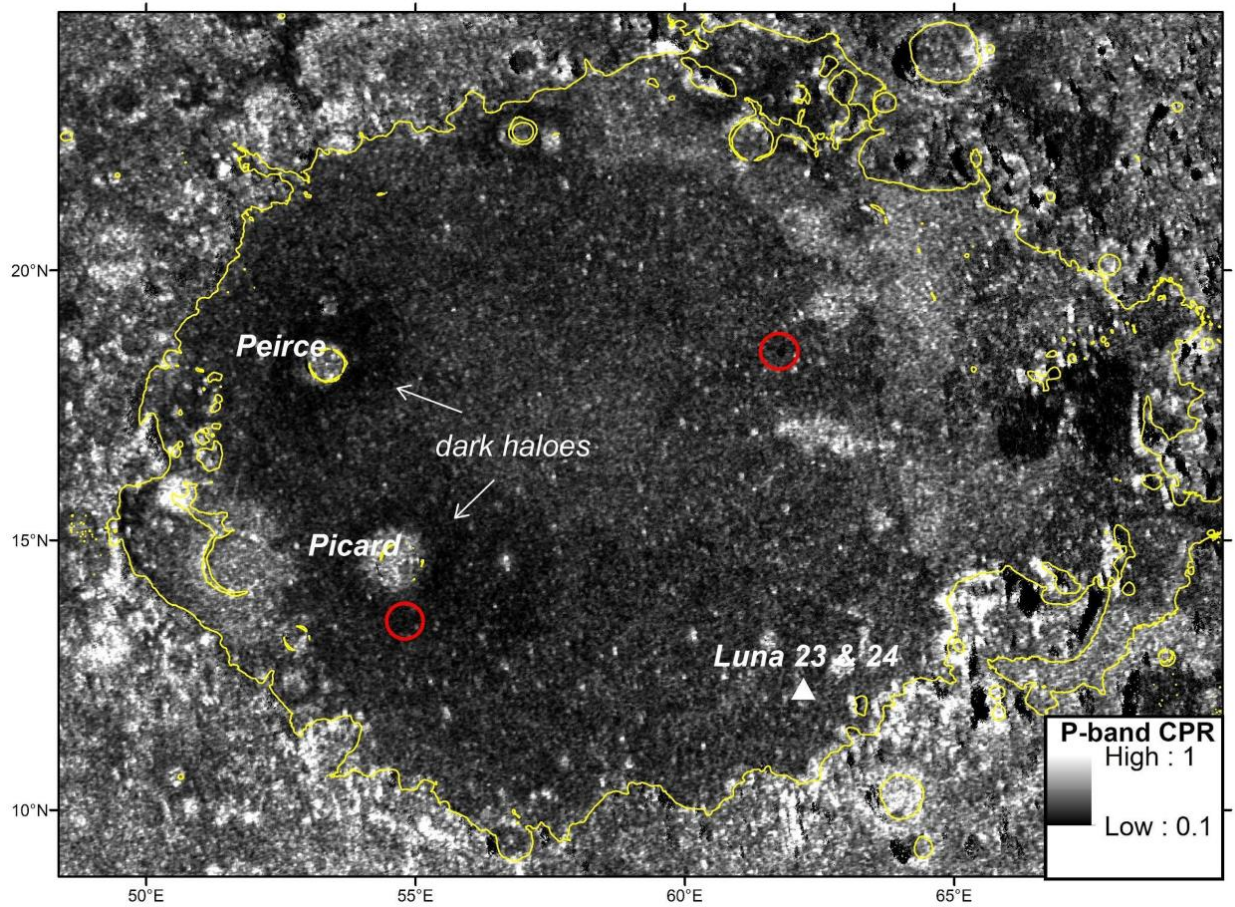

**Figure 11.** CPR image for the P-band radar returns obtained at the Arecibo Observatory (https://pds-geosciences.wustl.edu/missions/lunar_radar/index.htm). Yellow lines are -3-km elevation contours that roughly define the basin rims of Crisium. The white arrows point to the dark haloes around craters Peirce and Pickard. The red circles indicate the two candidate areas for further consideration for landing site selection. The white triangle indicates the landing sites of Luna 23 and 24.

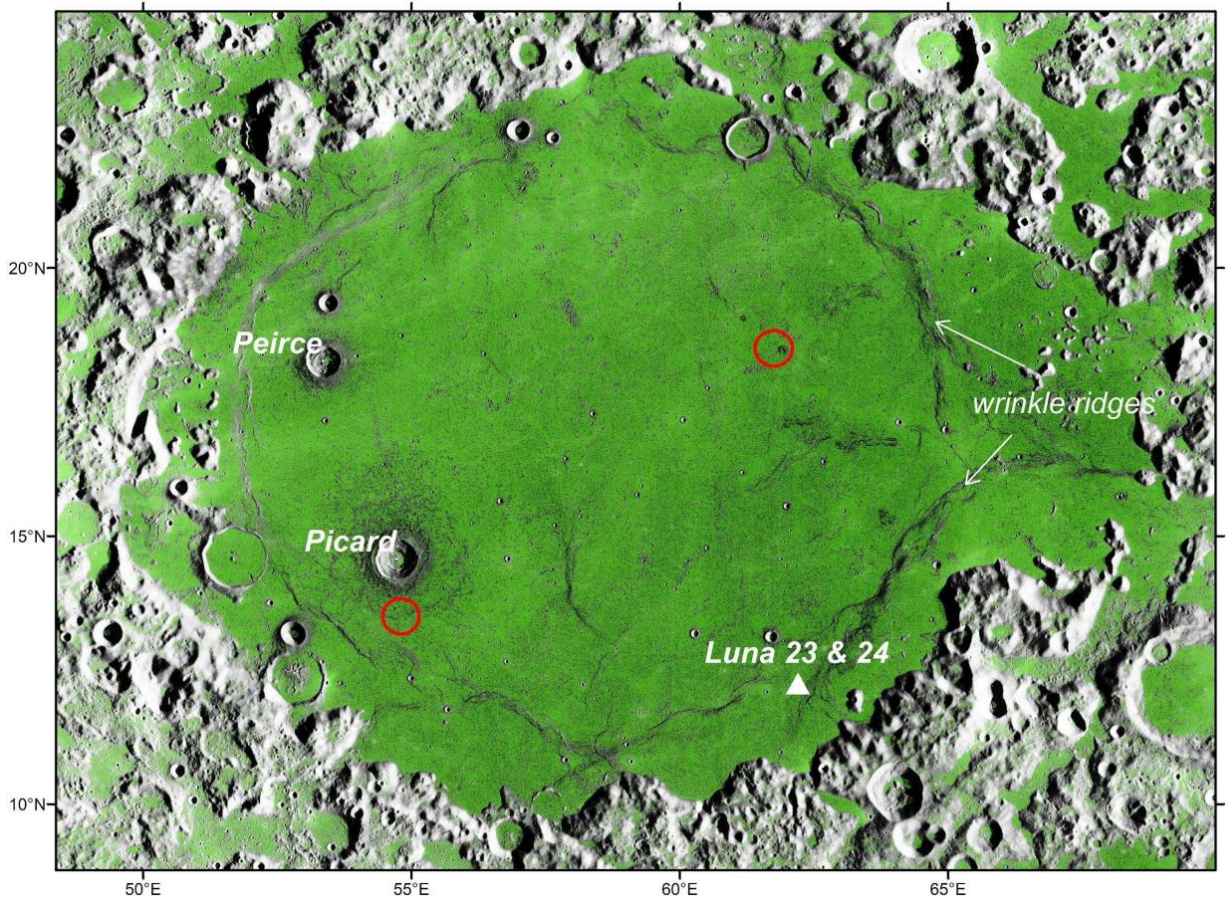

**Figure 12.** The areas of slope less than 4° are shown in green over the LRO-WAC mosaic, derived from the SLDEM (Barker et al., 2016) with 60 m spatial resolution. The white arrows point to some of the wrinkle ridges. The red circles indicate the two candidate areas for further consideration for landing site selection. The white triangle indicates the landing sites of Luna 23 and 24.

## 5. CONCLUSIONS

We have demonstrated the science driving network landing site selection for the LGN mission concept. These stations greatly improve upon the previous measurements made during Apollo and will provide key observations advancing knowledge of the lunar interior. While some of our science objectives may be accomplished with a single station, the full network, in the baseline or threshold configurations, is needed to provide sufficient science return. Optimization of network sites will at least double the rate of detection of lower mantle and core phases over the Apollo network. The wide distribution of stations also improves the density of seismic rays observed. NGLR improves the LLR network by pushing the stations towards the limbs with next generation reflectors. This long lived (>50 years) on-going program contributes to both geophysics and astrophysics with projected improvements indicating ~100x better accuracy. Heat flow and electromagnetics benefit from observations at multiple distinct locations representative of the lateral variability in internal structure. Once the network sites have been placed globally, locations within each landing site need to be analyzed across the spacecraft and instrument needs. We have demonstrated this methodology at the Crisium basin as one example. Future work will apply this methodology to each of the proposed LGN landing sites. These example landing sites will continue to be evaluated and analyzed throughout the development of the LGN mission. This will include detailed analyses conducted at each network location and updates will be evaluated across the primary and secondary mission requirements.

## Acknowledgements

D.C. acknowledges his research was carried out at the Jet Propulsion Laboratory, California Institute of Technology, under a contract with the National Aeronautics and Space Administration (80NM0018D0004) and at the University of Maryland, College Park under a contract with Notre Dame University (203769UMD). We acknowledge the use of imagery from Lunar QuickMap (https://quickmap.lroc.asu.edu), a collaboration between NASA, Arizona State University & Applied Coherent Technology Corp. Any use of trade, firm, or product names is for descriptive purposes only and does not imply endorsement by the U.S. Government.

**APPENDIX**

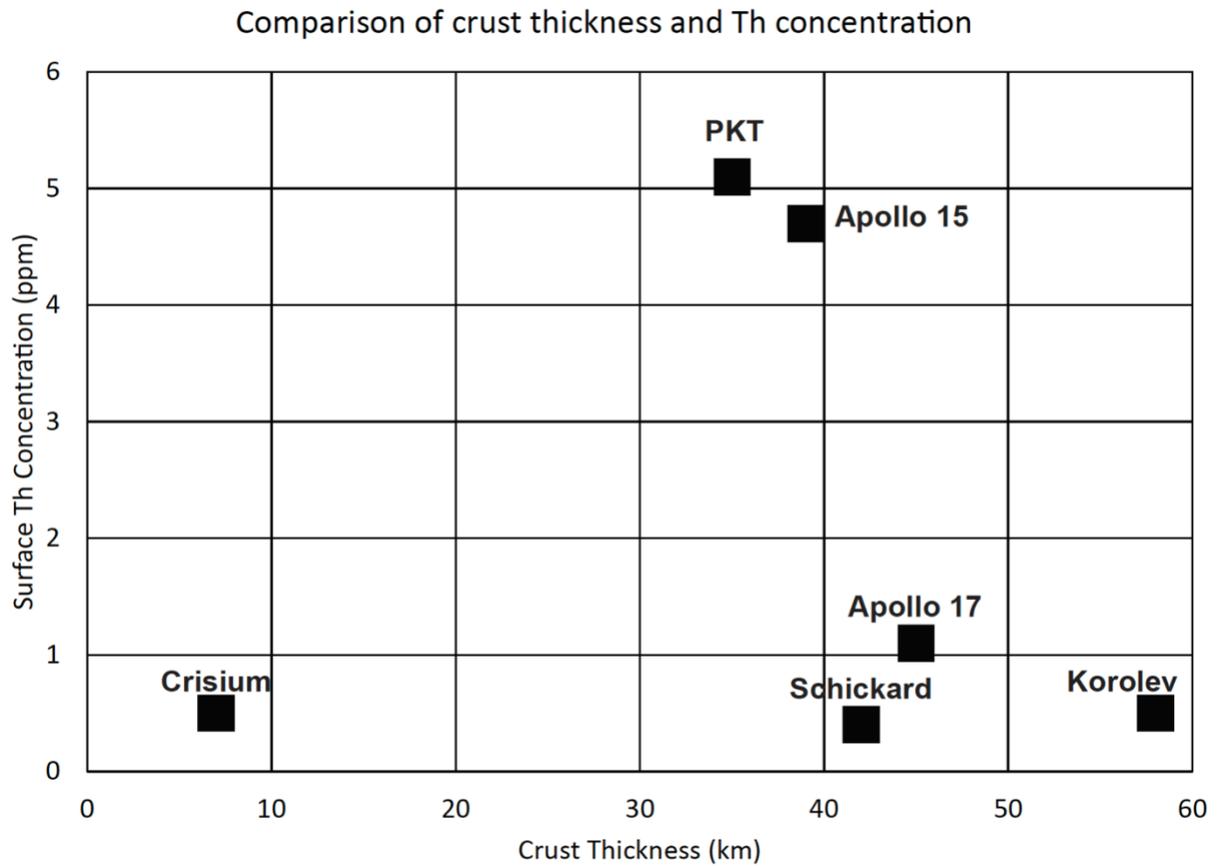

**Figure A1.** A scatter plot comparing the surface Thorium concentration (ppm) and crust thickness of the Apollo 15 and 17 sites and the proposed landing sites for the LGN mission.

LRO Quickmap links to LGN example landing sites:
- PKT: https://bit.ly/31LdM0e
- Schickard: https://bit.ly/3kyx5kF
- Crisium: https://bit.ly/3jtlv91
- Korolev: https://bit.ly/35DNjTc

**Table A1.** Acronym list

| | |
|---|---|
| ALSEP | Apollo Lunar Surface Experiment |
| APOLLO | Apache Point Observatory Lunar Laser-ranging Operation |
| CCRs | Cube Corner Retroreflectors |
| CLPS | Commercial Lunar Payload Services |
| CMB | Core Mantle Boundary |
| CPR | Circular Polarization Ratio |
| DEM | Digital Elevation Model |
| DMQ | Deep Moonquakes |
| DTM | Digital Terrain Model |
| ELGO | European Lunar Geophysical Observatory |
| EM | Electromagnetic |
| FHT | Feldspathic Highlands Terrain |
| GR | General Relativity |
| GRAIL | Gravity Recovery and Interior Laboratory |
| KREEP | Potassium, Rare Earth Elements, Phosphorus |
| LGN | Lunar Geophysical Network |
| LISTER | Lunar Instrumentation for Subsurface Thermal Exploration with Rapidity |
| LKFM | Low-K Fra Mauro |
| LLR | Lunar Laser Retroreflector |
| LLRO | Lunar Laser Ranging Observatories |
| LMAG | Lunar Magnetometer (Kaguya/SELENE) |
| LMS | Lunar Magnetotelluric Sounder |
| LOLA | Lunar Orbiter Laser Altimeter |
| LP | Lunar Prospector |
| LP | Long Period |
| LRO | Lunar Reconnaissance Orbiter |
| LROC | Lunar Reconnaissance Orbiter Camera |
| LSITP | Lunar Surface Instrument and Technology Payloads |
| MT | Magnetotelluric Method |
| NAC | Narrow Angle Camera |
| NGLR | Next Generation Lunar Retroreflectors |
| PcP | Core-Reflected P-wave |
| PKP | Core-Traversing P-wave |
| PKT | Procellarum KREEP Terrane |
| ScS | Core-Reflected Shear Wave |
| SEIS | Seismic Experiment for Interior Structure |
| SLR | Satellite Laser Ranging |
| SMQ | Shallow Moonquake |
| SP | Short Period |
| SSP | Silicon Seismic Package |
| SPA | South Pole-Aitken Basin |
| TF | Magnetic Transfer Function |
| VBB | Very Broad Band |
| VPREMOON | Very Preliminary REference MOON model |
| WAC | Wide Angle Camera |